\documentclass[aps,prd,reprint,twocolumn,superscriptaddress,longbibliography,nofootinbib,floatfix,showpacs]{revtex4-1}
\usepackage{epsfig}
\usepackage{amsmath,amssymb,amsfonts}
\usepackage{hyperref}
\usepackage{mathrsfs}
\usepackage{bbm}
\usepackage{slashed}
\usepackage{graphicx}
\usepackage{verbatim}

\usepackage{bm}

\usepackage[usenames]{xcolor}

\definecolor{darkgreen}{rgb}{0.2,0.6,0}

\newcommand{\be}{\begin{equation}}
\newcommand{\ee}{\end{equation}}
\newcommand{\bw}{\begin{widetext}}
\newcommand{\ew}{\end{widetext}}
\newcommand{\bi}{\begin{itemize}}
\newcommand{\ei}{\end{itemize}}
\newcommand{\ud}{\mathrm{d}}

\newcommand{\LCm}{{\scriptscriptstyle -}} \newcommand{\LCp}{{\scriptscriptstyle +}}
\newcommand{\LCpm}{{\scriptscriptstyle \pm}}

\newcommand{\LCperp}{{\scriptscriptstyle \perp}}

\newcommand{\pa}{\partial}
\newcommand{\bo}{\textbf}
\renewcommand{\Re}{\text{Re}}
\renewcommand{\Im}{\text{Im}}

\usepackage[T1]{fontenc} \usepackage[latin1]{inputenc}

\begin{document}

\title{Schwinger pair production in spacetime fields: Moir\'e patterns, Aharonov-Bohm phases and Sturm-Liouville eigenvalues}

\author{Gianluca Degli Esposti}
\email{g.degli-esposti@hzdr.de}
\affiliation{Helmholtz-Zentrum Dresden-Rossendorf, Bautzner Landstra{\ss}e 400, 01328 Dresden, Germany}

\affiliation{Institut f\"ur Theoretische Physik, 
Technische Universit\"at Dresden, 01062 Dresden, Germany}

\author{Greger Torgrimsson}
\email{greger.torgrimsson@umu.se}
\affiliation{Department of Physics, Ume{\aa} University, SE-901 87 Ume{\aa}, Sweden}

\begin{abstract}

We use a worldline-instanton formalism to study the momentum spectrum of Schwinger pair production in spacetime fields with multiple stationary points.  
We show that the interference structure changes fundamentally when going from purely time-dependent to space-time-dependent fields. For example, it was known that two time-dependent pulses give interference if they are anti-parallel, i.e. $E_z(t)-E_z(t-\Delta t)$, but here we show that two spacetime pulses will typically give interference if they instead are parallel, i.e. $E_z(t,z)+E_z(t-\Delta t,z-\Delta z)$. We take into account the fact that the momenta of the electron, $p_z$, and of the positron, $p'_z$, are independent for $E_z(t,z)$ (it would be $p_z+p'_z=0$ for $E(t)$), and find a type of fields which give moir\'e patterns in the $p_z-p'_z$ plane. Depending on the separation of two pulses, we also find an Aharonov-Bohm phase. We also study complex momentum saddle points in order to obtain the integrated probability from the spectrum. Finally, we calculate an asymptotic expansion for the eigenvalues of the Sturm-Liouville equation that corresponds to the saddle-point approximation of the worldline path integral, use that expansion to compute the product of eigenvalues, and compare with the result obtained with the Gelfand-Yaglom method.

\end{abstract}
\maketitle

\section{Introduction}

As shown in~\cite{Sauter:1931,Schwinger:1951}, a strong electric background field makes the vacuum unstable and production of pairs of electrons and positrons is expected to occur. Although first predicted almost a century ago, there has so far been no experimental evidence of the Sauter-Schwinger mechanism. The probability of this process has a nonperturbative exponential scaling\footnote{We absorb the electric charge into the field $eE \to E$ and use units where $c = \hbar = m_e = 1$, where in particular $E = 1$ corresponds to the Schwinger limit $E_S \simeq 10^{18}$ V/m.}
\be\label{eq:CFscaling}
\mathbb P \sim e^{-\frac{\pi}{E}} \;,
\ee
so pair production is highly suppressed for field strengths much smaller than the Schwinger limit $E \ll 1$.

Generalizing the constant-field result~\eqref{eq:CFscaling} to inhomogenous fields can be challenging. For 1D fields, $E_z(t)$, $E_z(z)$ or $E_z(t+z)$, one can use either Wentzel-Kramers-Brillouin (WKB) approximations~\cite{Brezin:1970xf,Popov:1972,Popov:2005,Kim:2000un,Kim:2007pm,Dumlu:2010ua,Dumlu:2011rr,Strobel:2013vza} or the worldline\footnote{The worldline formalism is often used without instantons, see reviews~\cite{Schubert:2001he,Edwards:2019eby}.} instanton formalism~\cite{Affleck:1981bma,Dunne:2005sx,Dunne:2006st,Dumlu:2011cc,ellipticalInstanton,Ilderton:2015qda}. 
However, while WKB works well for 1D fields, it is challenging to extend it to multidimensional fields (see~\cite{Kohlfurst:2021skr} for recent progress). The worldline instanton method, on the other hand, has now been shown to work also for multidimensional fields~\cite{Dunne:2006ur,Schneider:2018huk}. 
In~\cite{Schneider:2018huk}, the pair production probability was obtained numerically using discretized instantons\footnote{Discretized instantons have also been used in~\cite{Gould:2017fve}.} for an exact 4D solution to Maxwell's equation. However, most studies on worldline instantons have so far been for closed worldlines, which only\footnote{There is a trick that allows one to find the spectrum from closed instantons, as shown in~\cite{Dumlu:2011cc}, but it works only for 1D fields.} give the total/integrated probability via the imaginary part of the effective action. In~\cite{DegliEsposti:2021its,DegliEsposti:2022yqw,DegliEsposti:2023qqu} we have shown how to use open worldlines to obtain the momentum spectrum.

Fields with a single peak lead to a spectrum which also has a single peak. But, already for a purely time-dependent field, $E(t)$, if the field has several peaks then there can be rich interference effects in the spectrum~\cite{Dumlu:2010ua,Dumlu:2011rr,Hebenstreit:2009km,Dumlu:2010vv,Akkermans:2011yn}, and the precise pulse shape can have a strong effect already for time-dependent fields. Being able to make quantitative predictions of the spectra in realistic field configurations is not just of theoretical interest, it will also allow one to distinguish the desired signal from background processes in experiments. 

In this paper we generalize the open-instanton approach~\cite{DegliEsposti:2021its,DegliEsposti:2022yqw,DegliEsposti:2023qqu} to fields with multiple peaks. We show that going from a time-dependent field $E(t)$ to a field which also depends on space, $E_z(t,z)$, changes the interference structure fundamentally:

\bi
\item The condition for when a field gives interference changes completely because for $E(t)$ a particle created at an earlier peak necessarily goes through later peaks, while for $E(t,z)$ a particle created at one peak will in the generic case miss a later peak. Our instanton approach, especially when combined with a locally-constant-field (LCF) approximation, gives a simple way to predict the condition for interference.

\item Since the momenta of the electron and positron, $p_3$ and $p'_3$, are independent for $E(t,z)$, i.e. $p_3+p'_3\ne0$, we can have genuinely 2D interference patterns in the $p_3-p'_3$ plane, e.g. moir\'e patterns. Note that fully numerical approaches such as the Wigner formalism~\cite{Hebenstreit:2011wk,Kohlfurst:2015zxi,Ababekri:2019dkl,Aleksandrov:2019ddt,Kohlfurst:2021skr} have been applied to $E(t,z)$ fields but focus on a single momentum variable rather than both $p_3$ and $p'_3$. Thus, as far as we are aware, these moir\'e patterns have not been seen before.

\ei

Our instanton formalism also allows us to see Aharonov-Bohm phases for 2D fields with more than one peak.

In the application of WKB to $E(t)$ fields with a single peak, \cite{Popov:1972,Popov:2005} expanded the exponential part of the spectrum around the momentum saddle point,
\be\label{Pp3d}
\mathbb{P}(p_3)\propto\exp\left\{-\mathcal{A}-\frac{p_3^2}{d^2}\right\} \;,
\ee
where $d$ gives the width of the spectrum. However, we are not aware of such an expansion for $E(t)$ fields with multiple peaks. Perhaps the reason is due to the fact that for $E(t)$ it is relatively easy to divide the relevant interval in $p_3$ into a grid and then directly evaluate the WKB approximation at each gridpoint separately, i.e. without expanding around any momentum point; see e.g.~\cite{Dumlu:2010ua,Dumlu:2011rr}. However, when considering 2D fields $E(t,z)$, we have found it to be very useful to expand around momentum saddle points. This is true already for fields with a single peak~\cite{DegliEsposti:2022yqw,DegliEsposti:2023qqu}. Generalizing from $E(t)$ to $E(t,z)$ leads to more independent widths, because the electron and positron momenta are now independent, but that still means a huge reduction of the number of things to compute compared to a grid-based approach, and reducing the spectrum to a small number of quantities makes the computations much faster and also allows us to more easily grasp the dependence on the field parameters, e.g. by plotting $d(\gamma)$ with $\gamma=\omega/E_0$, where $\omega$ is a characteristic frequency scale and $E_0$ is the maximum field strength. Now when we turn to fields with multiple peaks, there are more than one instanton for each value of the momenta, and we do not just have widths, each term in the probability amplitude also has a wave vector ${\bm\alpha}$ giving oscillations in the $p_3-p'_3$ plane. Our results for ${\bm\alpha}$ allow one to design interference patterns by suitable choices of superpositions of multiple $E(t,z)$ pulses, e.g. to form moir\'e patterns. 

We will refer to this expansion around momentum saddle points as the quadratic approximation. We will show that it agrees well with the results from the grid approach, except near regions in momentum space where particles created by one field peak are scattered by another peak.

\section{Amplitude}

Our starting point is the Lehmann-Symanzik-Zimmermann (LSZ) reduction formula~\cite{ItzyksonZuber,Barut:1989mc} ($px=p_\mu x^\mu$, $g_{\mu\nu}=\text{diag}(1,-1,-1,-1)$),
\be\label{eq:LSZ3pair}
\begin{split}
M=\lim_{t_\LCpm\to\infty}\int \ud^3x_\LCp\ud^3 x_\LCm e^{ipx_\LCp+ip'x_\LCm}
\bar{u}\gamma^0S(x_\LCp,x_\LCm)\gamma^0 v \;,
\end{split}
\ee
where $u$ and $v$ are the free fermion spinors and $S$ is the fermion propagator in a background field. 
The reason we can use free fermion states is because both the electric field and the gauge potential vanish asymptotically, $A_\mu(t\to\infty,z\to\pm\infty)=0$; see~\cite{DegliEsposti:2022yqw,DegliEsposti:2021its}.
In the worldline formalism $S$ can be expressed as a single-particle path integral
\be\label{propagatorWorldline}
\begin{split}
&S(x_\LCp,x_\LCm)=(i\slashed{\partial}_{x_\LCp}-\slashed{A}(x_\LCp)+1)\int_0^\infty\frac{\ud T}{2}\int\limits_{q(0)=x_\LCm}^{q(1)=x_\LCp}\mathcal{D}q\,\mathcal{P}\\
&\times\exp\left\{-i\left[\frac{T}{2}+\int_0^1\!\ud\tau\left(\frac{\dot{q}^2}{2T}+A\dot{q}+\frac{T}{4}\sigma^{\mu\nu}F_{\mu\nu}\right)\right]\right\} \;.
\end{split}
\ee
The variable $T$ can be interpreted as total proper time, $\tau$ is the normalized proper time, $\mathcal P$ is proper-time ordering, and $\sigma^{\mu\nu}=\frac{i}{2}[\gamma^\mu,\gamma^\nu]$. 

Proper-time derivatives are denoted as $\dot{q}=\ud q/\ud \tau$ or $q'=\ud q/\ud u$, where $u=T(\tau-\sigma)$ and $\sigma\sim1/2$ is chosen so that the instanton is inside the field at $u \sim 0$. We use $\tau$ to derive the equations of motion, but to solve them we use $u$. 

The calculation for the contribution from a single instanton is the same as in~\cite{DegliEsposti:2022yqw,DegliEsposti:2023qqu}, except that now, when the amplitude contains several terms, we have to calculate the relative phases of each term. 

The integrals in~\eqref{eq:LSZ3pair} are nontrivial and cannot be performed exactly. However, by rescaling the variables as
\be
q^\mu \to \frac{q^\mu}{E_0} \qquad T \to \frac{T}{E_0}
\ee
one notices that all terms in the exponent, except the spin term, are of order $\mathcal O(1/E_0)$, so for $E_0\ll1$ we can perform the integrals with the saddle-point method. The spin term, $\sigma^{\mu \nu}F_{\mu \nu}$ is $\mathcal O(1)$ and therefore does not affect the saddle-point equations. 

The saddle points are determined from the equations obtained by setting to zero the variations of the $\mathcal{O}(1/E_0)$ part of the exponent with respect to $q_\mu$, $z_\LCpm$ and $T$. We find the Lorentz-force equation 
\be\label{LorentzEq}
t''=E(t,z)z' \qquad z''=E(t,z)t' \; ,
\ee
together with conditions for the asymptotic momenta
\be\label{eq:SPE}
z'(u_1) = -p_3 \qquad z'(u_0) =p'_3 \;,
\ee
where $u_0\to-\infty$ and $u_1\to+\infty$,
and an on-shell condition
\be
t^{\prime2}(u)-z^{\prime2}(u)=m_\LCperp^2 \;,
\ee
where $m_\LCperp=\sqrt{1+p_\LCperp^2}$ and $p_\LCperp$ is the momentum components transverse to the field. We focus on $p_\LCperp=0$ since the spectrum is peaked at $p_\LCperp=0$. 
Note that we now use $q^\mu(u)$ to denote the saddle point, i.e. the instanton, rather than the path-integration variable.

In terms of the variable $u$, the field is nonzero (or non-negligible) only in a bounded interval $(\tilde u_0, \tilde u_1)$ around $u=0$. Outside this interval, since the field vanishes, the instantons satisfy $t'' = z'' = 0$ and are therefore straight lines. In these asymptotic regions the instantons describe free propagating particles with momenta~\eqref{eq:SPE}.
In the derivation of~\eqref{eq:SPE} we had $u_0 = -T\sigma$ and $u_1 = T(1-\sigma)$, which correspond to the initial and final points $\tau = 0$ and $\tau = 1$. However, we do not need to determine $T$, we just need to know that $\text{Re}(T)\gg1$, because this means we can just choose some non-unique values of $-u_0$ and $u_1$ as long as they are large enough so that the instanton has left the field and $t'$ and $z'$ are constant. Thus,
$T$ and $z_\LCpm$ do not appear in~\eqref{LorentzEq} and~\eqref{eq:SPE}. $T$ and $z_\LCpm$ only played a nontrivial role in the derivation but not in the evaluation of the general formulas. 

Since we have a tunneling process, the instantons must be complex at least in some propertime interval. In the usual closed-instanton formalism, there is no connection to the real trajectories that the produced particles are expected to follow asymptotically after leaving the field. In~\cite{DegliEsposti:2023qqu} we found a complex proper-time contour that splits the instantons into a formation region, where the instanton is complex and where the ``creation happens'', and an acceleration region, where the instanton is real but still inside the field and hence follows the trajectory of a real (anti-)particle. We have referred to this as the ``physical'' contour. By ``physical'' we do not mean ``actual/correct'', because there is no such thing as ``the actual/correct'' trajectory of these particles. One has a great deal of freedom to deform this contour, as long as one does not pass poles or branch points. Different contours give different complex trajectories. One can even choose contours for which $t$ and $z$ never become real, i.e. $\text{Im }q^\mu(\pm\infty)\ne0$ even though $\text{Im }q_\mu'(\pm\infty)=0$; see e.g. the ``tilted'' contours in~\cite{DegliEsposti:2022yqw}. All contours which are continuously deformable to the physical contour give exactly the same results for all the observables, i.e. the asymptotic momenta and spins. Thus, the physical contour is not unique or necessary. However, it is often convenient for calculations, because it roughly minimizes the interval over which the instanton is complex.  

As shown in~\cite{Dumlu:2011cc}, if the field has several peaks, then there are multiple instantons contributing to the path integral,
\be\label{eq:PIsaddles}
\int \mathcal Dq \, e^{i \mathcal I[q]} \sim \sum_j e^{i \mathcal I[q^{(j)}]}  \;,
\ee
causing interference effects. For time-dependent fields, considered in previous works, one has $p'_3=-p_3$ and the interference patterns are therefore 1D too. For $E_3(t,z)$, the momentum along the $z$-axis is no longer conserved, $p'_3\ne-p_3$, and we will show that the interference pattern can either be stripes in the $p_3-p'_3$ plane, a simple generalization of the 1D patterns, or genuinely 2D, such as moir\'e patterns.

From the transverse components of the integrals in~\eqref{eq:LSZ3pair} we obtain a momentum conserving delta function
\be
(2\pi)^2 \delta_\LCperp(p+p') \;,
\ee
where $p_\LCperp=\{p_1, p_2\}$.

We showed in~\cite{DegliEsposti:2021its} that the exponential part of the probability amplitude, $M\propto e^{\psi}$, can be expressed as
\be\label{eq:psiIntegral}
\psi := i\int \ud u \, q^\mu \pa_\mu A_\nu \frac{\ud q^\nu}{\ud u} \;,
\ee
where the instanton $q$ depends on the momenta through the boundary conditions~\eqref{eq:SPE}. The instantons (and hence the integrand) are analytic functions almost everywhere apart from branch points, so we can deform the $u$-contour as long as we do not cross such points~\cite{DegliEsposti:2023qqu}. 

As to the prefactor, in the asymptotic limit $t_\LCpm \to \infty$ both the path integral and the ordinary integrals produce contributions proportional to some power of $t_\LCpm$, so it is convenient to extract such dependences analytically and cancel them out before doing any numerical computation. We showed how this can be done in~\cite{DegliEsposti:2022yqw}. The last contribution, namely the spin structure, is trivial for these 2D electric fields. However, now that more than one instanton contributes to the amplitude, we need to be careful about the relative signs due to the spin factor (see Appendix~\ref{App:Spin}).

Using the results from~\cite{DegliEsposti:2023qqu,DegliEsposti:2022yqw}, the contribution from the instanton $j$ to the probability amplitude is given by
\be\label{eq:Mj}
M_j=\varepsilon \, \delta_{ss'} (2\pi)^2 \delta_\LCperp(p+p') \sqrt{\frac{2\pi m_\perp^2}{p_0 p'_0 \, h(\tilde u_1)}}  \; e^{\psi_j} \;,
\ee
where $\psi_j$ is given by~\eqref{eq:psiIntegral}, $\varepsilon = \pm 1$ is a relative sign coming from the spin factor, and $h(\tilde u_1) = \eta'(\tilde u_1)$ comes from the functional determinant of the worldline path integral. We found a simple way to compute this factor by solving
\be\label{etaEq}
\eta'' = (E^2 + \nabla E \cdot\{z',t'\}) \eta \;,
\ee
where $\nabla E=\{\partial_t E,\partial_z E\}$, $E(t,z)$ is evaluated at the instanton solution, with initial conditions
\be
\eta(\tilde u_0) = 1, \quad \eta'(\tilde u_0) = 0 \;.
\ee
Eq.~\eqref{etaEq} comes from expanding the Lorentz-force equation to linear order around the instanton solution.

\section{Spectrum}\label{sec:Spectrum}

We set $p_\LCperp = 0$, which maximizes the probability, and focus on the longitudinal momentum components $\Pi=(p_3, p'_3)$, including in the spectrum the factors of $2\pi$ in the momentum integral. With $M_j$ as in~\eqref{eq:Mj}, we define the spectrum by summing over spins and integrating over $p'_\LCperp$. In the saddle-point approximation, the individual terms in the amplitude scale with respect to the field strength $E_0$ as
\be
M(\Pi) = E_0^a \, C(\Pi) \, e^{\frac{\psi(\Pi)}{E_0}}
\ee
where $a$ is some real number that depends on the dimensionality of the field, and $C$ and $\psi$ are some $E_0$-independent functions. Usually we do not write the $E_0$ factors explicitly and absorb them into $C$ and $\psi$. To compute the spectrum, however, we need to find the power $a$ in the prefactor. We get a factor of $1/\sqrt{E_0}$ from each ordinary integral, i.e. $T, z_\LCpm$, and two out of the four perpendicular integrals $x^\LCperp_\LCpm$, since two of them give the delta function in~\eqref{eq:Mj}. Furthermore, we get a factor of $\sqrt{E_0}$ from each path integral variable, therefore in total $a=-1/2$. 

The spectrum is thus given by
\be
\begin{split}
    \mathbb P(\Pi) 
    &:= \frac{1}{E_0}\sum_{ss'}\frac{1}{(2\pi)^5}\int \frac{\ud p'_\LCperp}{(2\pi)} \Bigr| \sum_j M_j \Bigr|^2 \\
    &=V_{\LCperp} \frac{2}{(2\pi)^3 \, p_0 p'_0 \, E_0} \left|\sum_j \varepsilon \frac{e^{\psi_j(\Pi)/E_0}}{\sqrt{h_j(\tilde u_1)}} \right|^2
\end{split}
\ee
where $h_j(\tilde u_1)$ is the solution to~\eqref{etaEq} calculated with the $j$-th instanton and $V_\LCperp = V_1 V_2$.

We have two approaches for computing the spectrum. One is to construct a grid of points in $(p_3,p_3')$ and compute instantons at each point. In cases of interference, there are more than one instanton for each value of $(p_3,p_3')$. Given the much larger number of instantons that need to be computed, this approach is significantly slower than the following ``quadratic-approximation'' approach.

Each amplitude term has the form
\be\label{MCE}
M(\Pi) = C(\Pi) \, e^{\psi(\Pi)} \;,
\ee
where $\psi \sim 1/E_0$. For $E_0 \ll 1$ the exponential is highly suppressed when the momentum is far away from the maximum $\Pi_s$ of $\Re \, \psi$, so only values of $\Pi$ which are close to $\Pi_s$ are important. In the important region we can therefore Taylor expand the exponential part of the spectrum up to second order in $|\Pi-\Pi_s|$,
\be\label{eq:expExpanded}
\begin{split}
    \psi(\Pi) \simeq& \; \psi(\Pi_s) + \nabla \psi(\Pi_s) \cdot (\bm \Pi - \bm \Pi_s) \\
    &+ \frac{1}{2}(\bm \Pi - \bm \Pi_s) \cdot \nabla \otimes \nabla \psi (\Pi_s) \cdot (\bm \Pi - \bm \Pi_s) \\
    =& \; i\phi -\frac{\mathcal A}{2} + i\bm \alpha \cdot (\bm \Pi - \bm \Pi_s) \\
    &+ \frac{1}{2}(\bm \Pi - \bm \Pi_s) \cdot \left(-\bo d^{-2} + i\bm \beta \right) \cdot (\bm \Pi - \bm \Pi_s) \;,
\end{split}
\ee
where
\be\label{eq:GeneralParameters}
\begin{split}
    \{-\mathcal A/2, \phi \} &= \{\text{Re},\text{Im}\} \, \psi(\Pi_s) \\
    \{\bm \alpha,\bm \rho\} &= \{\Im \nabla \psi(\Pi_s),\Re \nabla \psi(\Pi_s)\} \\
    \{-\bo d^{-2},\bm \beta\} &= \{\text{Re},\text{Im}\} \, \nabla \otimes \nabla \psi (\Pi_s) \; .
\end{split}
\ee
Note that there is no real linear term in~\eqref{eq:expExpanded} because $\Pi_s$ is by definition the saddle point of $\Re \, \psi$, i.e. ${\bm\rho}=0$.   
We will show below that the momentum saddle point for the entire $\psi$ is in general complex. This complex saddle point is relevant for integrating over the momenta to find the total probability, but for the spectrum it makes, for experimental reasons, more sense to expand around a real value of the momenta. $\mathcal{A}$ gives the overall exponential suppression of the entire spectrum. $\phi$ includes Aharonov-Bohm phases. ${\bm\alpha}$ describes both the frequency and the direction of oscillations in the spectrum. ${\bf d}^{-2}$ gives the widths of the peaks in the spectrum.
In the pre-exponential factor in~\eqref{MCE} we can simply set $\Pi=\Pi_s$.

To compute the parameters~\eqref{eq:GeneralParameters} we take derivatives of the exponent in the initial form~\eqref{eq:LSZ3pair}
\be\label{eq:PsiInitial}
\psi = ipx_\LCp + ip'x_\LCm - \frac{iT}{2}-i\int_0^1 d\tau \, \frac{\dot q^2}{2T} + Aq
\ee
and evaluate it $\Pi=\Pi_s$. Because of symmetries, it is convenient to change the momentum variables from $p_3, p'_3$ to
\be\label{p3p3PDelta}
p_3 = -P + \frac{\Delta}{2} \qquad p'_3 = P + \frac{\Delta}{2} \; .
\ee

The exponent~\eqref{eq:PsiInitial} depends on $\Pi$ explicitly but also implicitly as the saddle points $z_\LCpm$, $T$, and $q^\mu$ are functions of $\Pi$. However, when differentiating with respect to $\Pi$, we can neglect the implicit dependence because the partial derivatives of $\psi$ with respect to these integration variables vanish, since they are evaluated at saddle points. The imaginary linear terms are therefore simply
\be\label{eq:alphas}
\begin{split}
    \alpha_\Delta &= \frac{1}{2} \Re \, \left[ z(u_1) +z(u_0) +\frac{p_3}{p_0}t(u_1) +\frac{p'_3}{p'_0}t(u_0) \right] \\
    \alpha_P &= \Re \, \left[ -z(u_1) +z(u_0) -\frac{p_3}{p_0}t(u_1) +\frac{p'_3}{p'_0}t(u_0) \right] \; .
\end{split}
\ee
Note that, once we have found the instanton, there are no additional differential equations to solve or integrals to compute, as~\eqref{eq:alphas} gives us ${\bm\alpha}$ by simply evaluating the instanton at asymptotic points.  

To compute the $\mathcal{O}([\Pi-\Pi_s]^2)$ terms, we expand the first-order derivatives (given by~\eqref{eq:alphas} without taking the real part) to linear order in $\delta P = P - P_s$ and $\delta\Delta = \Delta - \Delta_s$. To do so we need the variations of the instanton ($q^\mu \to q^\mu + \delta q^\mu$) due to $\delta P$ and $\delta\Delta$. These variations are obtained by expanding the Lorentz-force equation around $\delta P=\delta\Delta=0$. We find~\cite{DegliEsposti:2023qqu,DegliEsposti:2022yqw} that $\delta t$ and $\delta z$ satisfy
\be\label{eq:deltaqEqs}
\begin{split}
\delta t''&=E\delta z'+\nabla E\cdot\{\delta t,\delta z\}z' \\
\delta z''&=E\delta t'+\nabla E\cdot\{\delta t,\delta z\}t' \;,
\end{split}
\ee
where $\nabla E=\{\partial_t E,\partial_z E\}$. We need two solutions, $\delta q_\Delta$ and $\delta q_P$. $\delta q_\Delta$ is due to a variation in $\delta\Delta$ with $\delta P=0$, and vice versa for $\delta q_P$. By varying the asymptotic momenta we find the following boundary conditions,
\be\label{eq:dDeltaCond}
\delta z_\Delta'(\pm\infty)=\mp\frac{1}{2} \qquad \delta t_\Delta'(\pm\infty)= -\frac{P}{2p_0}
\ee
and
\be\label{eq:dPCond}
\delta z_P'(\pm\infty)=1 \qquad \delta t_P'(\pm\infty)=\pm\frac{P}{p_0} \; .
\ee
At first glance, from the boundary conditions~\eqref{eq:dDeltaCond} and~\eqref{eq:dPCond} it looks like we must use the shooting method to find $\delta q_P^\mu$ and $\delta q_\Delta^\mu$ as well as the instantons $q^\mu$. This would lead to relatively slow numerics, because in the shooting method one has to solve a differential equation several times with different initial conditions until some asymptotic condition is satisfied. However, as we showed in~\cite{DegliEsposti:2023fbv}, the $\delta q$'s can be expressed in terms of a basis set of variations $\delta q_{[j]}$ with simple initial conditions at $u=0$
\be
\delta t_{[1]}(0) = \delta z_{[2]}(0) = \delta t'_{[3]}(0) = \delta z'_{[4]}(0) = 1
\ee
and all others being zero. The reason we can avoid the shooting method is basically because~\eqref{eq:deltaqEqs} is a linear and homogeneous equation, so after we have found the four basis solutions $\delta q_{[j]}$ we can find the coefficients in the superposition by simply solving an algebraic equation. As described in the next section, the middle of the instanton is a turning point, where $t'(0)=0$ and $z'(0)=\pm i$, which implies 
\be
\delta t_P'(0)=\delta z_P'(0)=\delta t_\Delta'(0)=\delta z_\Delta'(0)=0 \;.
\ee
So here we actually only need two of the four basis solutions, 
\be
\begin{split}
\delta q^\mu_P(u)&=a_P \delta q_{[1]}^\mu(u)+b_P \delta q_{[2]}^\mu(u)\\
\delta q^\mu_\Delta(u)&=a_\Delta \delta q_{[1]}^\mu(u)+b_\Delta \delta q_{[2]}^\mu(u) \;,
\end{split}
\ee
where
\be
\begin{split}
a_P&=\frac{\delta z_{[2]}'(\infty)-\delta z_{[2]}'(-\infty)}{\delta z_{[1]}'(-\infty)\delta z_{[2]}'(\infty)-\delta z_{[1]}'(\infty)\delta z_{[2]}'(-\infty)}\\
b_P&=-\frac{\delta z_{[1]}'(\infty)-\delta z_{[1]}'(-\infty)}{\delta z_{[1]}'(-\infty)\delta z_{[2]}'(\infty)-\delta z_{[1]}'(\infty)\delta z_{[2]}'(-\infty)}\\
a_\Delta&=\frac{1}{2}\frac{\delta z_{[2]}'(\infty)+\delta z_{[2]}'(-\infty)}{\delta z_{[1]}'(-\infty)\delta z_{[2]}'(\infty)-\delta z_{[1]}'(\infty)\delta z_{[2]}'(-\infty)}\\
b_\Delta&=-\frac{1}{2}\frac{\delta z_{[1]}'(\infty)+\delta z_{[1]}'(-\infty)}{\delta z_{[1]}'(-\infty)\delta z_{[2]}'(\infty)-\delta z_{[1]}'(\infty)\delta z_{[2]}'(-\infty)} \;.
\end{split}
\ee

Letting $X_{ij}$ be the second variation/derivative of the exponent, with $i,j \in \{P, \Delta\}$, we have
\be\label{d2andbetafromX}
d_{ij}^{-2} = -\Re \, X_{ij} \qquad \beta_{ij} = \Im \, X_{ij}
\ee
where, as shown in~\cite{DegliEsposti:2023fbv},
\be
\begin{split}
X_{PP}=i&\bigg[-\delta z_P(\infty)-\frac{p_3}{p_0}\delta t_P(\infty)+\frac{t(\infty)}{p_0^3}\\
&+\delta z_P(-\infty)+\frac{p'_3}{p'_0}\delta t_P(-\infty)+\frac{t(-\infty)}{p_0^{\prime3}}\bigg]    
\end{split}
\ee
\be
\begin{split}
X_{\Delta P}=&\frac{i}{2}\bigg[\delta z_P(\infty)+\frac{p_3}{p_0}\delta t_P(\infty)-\frac{t(\infty)}{p_0^3}\\
&+\delta z_P(-\infty)+\frac{p'_3}{p'_0}\delta t_P(-\infty)+\frac{t(-\infty)}{p_0^{\prime3}}\bigg] \\
=&i\bigg[-\delta z_\Delta(\infty)-\frac{p_3}{p_0}\delta t_\Delta(\infty)-\frac{t(\infty)}{2p_0^3}\\
&+\delta z_\Delta(-\infty)+\frac{p'_3}{p'_0}\delta t_\Delta(-\infty)+\frac{t(-\infty)}{2p_0^{\prime3}}\bigg]
\end{split}
\ee 
\be
\begin{split}
X_{\Delta\Delta}=\frac{i}{2}&\bigg[\delta z_\Delta(\infty)+\frac{p_3}{p_0}\delta t_\Delta(\infty)+\frac{t(\infty)}{2p_0^3}\\
+&\delta z_\Delta(-\infty)+\frac{p'_3}{p'_0}\delta t_\Delta(-\infty)+\frac{t(-\infty)}{2p_0^{\prime3}}\bigg]
\end{split}
\ee
where the momenta $p_3, p'_3$ denote their saddle-point values.
From now on, $q(u)$ denotes the instanton at the momentum saddle point $\Pi_s$. 

In essence, we have reduced the spectrum to a relatively small number of parameters which can be obtained either from an integral~\eqref{eq:psiIntegral} or differential equation~\eqref{eq:deltaqEqs} involving the instantons. If there is only one instanton, taking the modulus squared removes all the imaginary terms and we are left with
\be
\mathbb P(\Pi) = |C(\Pi_s)|^2 \, e^{-\mathcal A - (\bm \Pi - \bm \Pi_s) \cdot \bo d^{-2} \cdot (\bm \Pi - \bm \Pi_s)}
\ee
as in~\cite{DegliEsposti:2022yqw,DegliEsposti:2023qqu}. But for fields with multiple peaks the imaginary contributions are precisely those giving rise to interference effects in the momentum
\be\label{eq:specTotal}
\begin{split}
    \mathbb P(\Pi) =& \sum_i |C_i|^2 \, e^{2 \Re \, \psi_i(\Pi)}\\
    &+\sum_{i<j} 2\Re \left[ C_i C_j^* \, e^{\psi_i(\Pi) + \psi_j(\Pi)^*} \right] \; .
\end{split}
\ee

\begin{figure*}
    \includegraphics[width=0.49\linewidth]{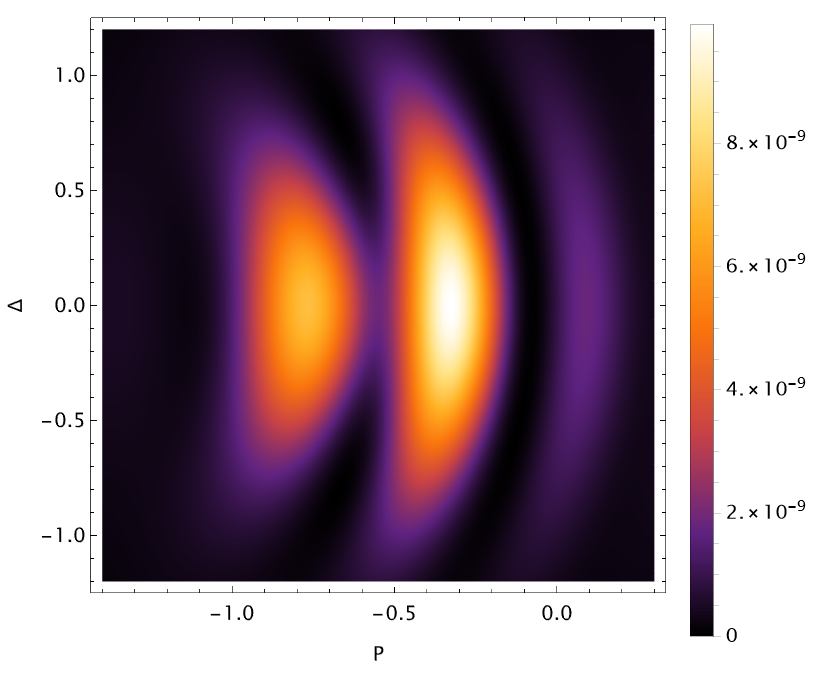} \hfill
    \includegraphics[width=0.49\linewidth]{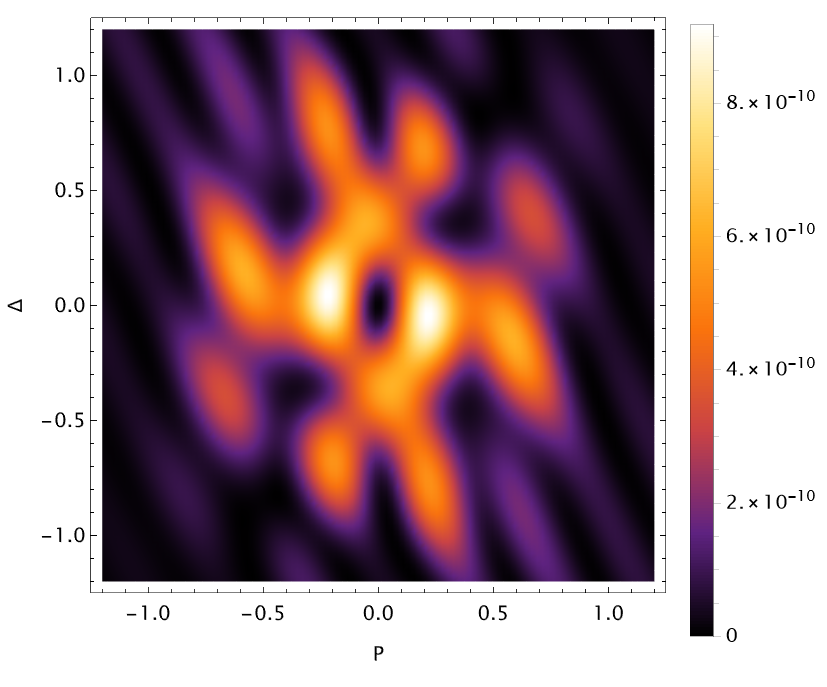}
    \caption{Spectra for the fields~\eqref{eq:field21} (left) and~\eqref{eq:field22} (right) with parameters $\gamma_t = \omega_t/E = 1$, $\gamma_z = \omega_z/E = 1$ and $E_0=0.2$. $P$ and $\Delta$ are related to the electron and positron momenta, $p_3$ and $p'_3$, as in~\eqref{p3p3PDelta}.}
    \label{fig:spectra1}
\end{figure*}

In Fig.~\ref{fig:spectra1} we can see two example spectra for the fields
\be\label{eq:field21}
E_3(t,z)/E_0 = -2\omega_t t \, e^{-(\omega_t t)^2 - (\omega_z z)^2}
\ee
with two time peaks (and hence two instantons), and
\be\label{eq:field22}
E_3(t,z)/E_0 = 4\omega_t t \, \omega_z z \, e^{-(\omega_t t)^2 - (\omega_z z)^2}
\ee
with two time and two space peaks (hence four instantons). 

As we point out in Appendix~\ref{App:Symmetric}, for symmetric fields such as~\eqref{eq:field21} we have $\alpha_\Delta=0$, so the interference is only along the $P$-axis. On the other hand, for a general field such as~\eqref{eq:field22}, the interference pattern has a richer structure as we can see from the plot on the right of Fig.~\ref{fig:spectra1}.

\section{Instantons}

To compute the instantons at a generic momentum value ($\Pi$ not necessarily equal to $\Pi_s$) we use the shooting method, varying the initial conditions at the turning point $u = 0$ until~\eqref{eq:SPE} are satisfied. $t'(0) = 0$ follows from the fact that $u = 0$ is a turning point, and $z'(0) = \varepsilon i$ with $\varepsilon = \pm 1$ from the on-shell condition $q'(0)^2 = 1$, so for a general field we are left with two unknown complex parameters $t(0)$ and $z(0)$.

In order to find $\Pi_s$, one could in principle find the instantons at several values of the momenta until the maximum of $\Re \, \psi(\Pi)$ is found. This procedure is time consuming because for every value of the momentum one has to use the shooting method to find the initial conditions and calculate an integral to find $\psi$. For the fields we consider here, $p_{\LCperp s} = 0$, so the only nontrivial saddle points are for the longitudinal components $p_{3s}$ and $p'_{3s}$. A simpler way to find $\Pi_s$ is to use the shooting method once with conditions
\be\label{eq:SPconstraints}
\begin{split}
    \frac{\pa \Re \, \psi}{\pa p_3} &= \Im\left[z(\infty) - \frac{ z'(\infty)}{t'(\infty)}t(\infty) \right] = 0 \\
    \frac{\pa \Re \, \psi}{\pa p'_3} &= \Im\left[ z(-\infty) - \frac{ z'(-\infty)}{t'(-\infty)}t(-\infty) \right] = 0 \;,
\end{split}
\ee
which gives directly the instantons at $\Pi_s$ without having to find any instantons for $\Pi\ne\Pi_s$. The value of $\Pi_s$ is then obtained by simply evaluating $z'(\pm\infty)$.

There is an apparent mismatch between the number of free parameters (four real, $t(0)$ and $z(0)$), and the constraints~\eqref{eq:SPconstraints} (two real). However, since the asymptotic momenta must be real, we have two additional constraints
\be\label{eq:RealMomentum}
\Im \, z'(\pm \infty) = 0 \; .
\ee

For symmetric fields, the additional symmetry of the instantons reduces the number of parameters that need to be varied in the shooting method, and hence speeds up the numerics. If the field is even with respect to $z$, $E_z(t,z) = E_z(t,-z)$, $t(u)$ and $z(u)$ are respectively even and odd for $\Pi=\Pi_s$, so $z(0) = 0$ and we therefore only need to vary $t(0)$. The number of constraints is also reduced to two, as we only need to consider~\eqref{eq:SPconstraints} and~\eqref{eq:RealMomentum} at $u = +\infty$ (or $-\infty$). The parameters ${\bm\alpha}$, etc are also simpler, as we show in Appendix~\ref{App:Symmetric}.

However, even for symmetric fields, the instantons are in general not symmetric for $\Pi\ne\Pi_s$. Indeed, if $\Delta \neq 0$ the asymptotic momenta $z'(\pm \infty)$ are different, i.e. there is no symmetry between the final electron and positron momenta, and then $z(u)$ is not odd. Therefore, finding instantons away from the saddle point is computationally more time consuming because one would have to vary twice as many parameters.

\begin{figure*}
    \includegraphics[width=\linewidth]{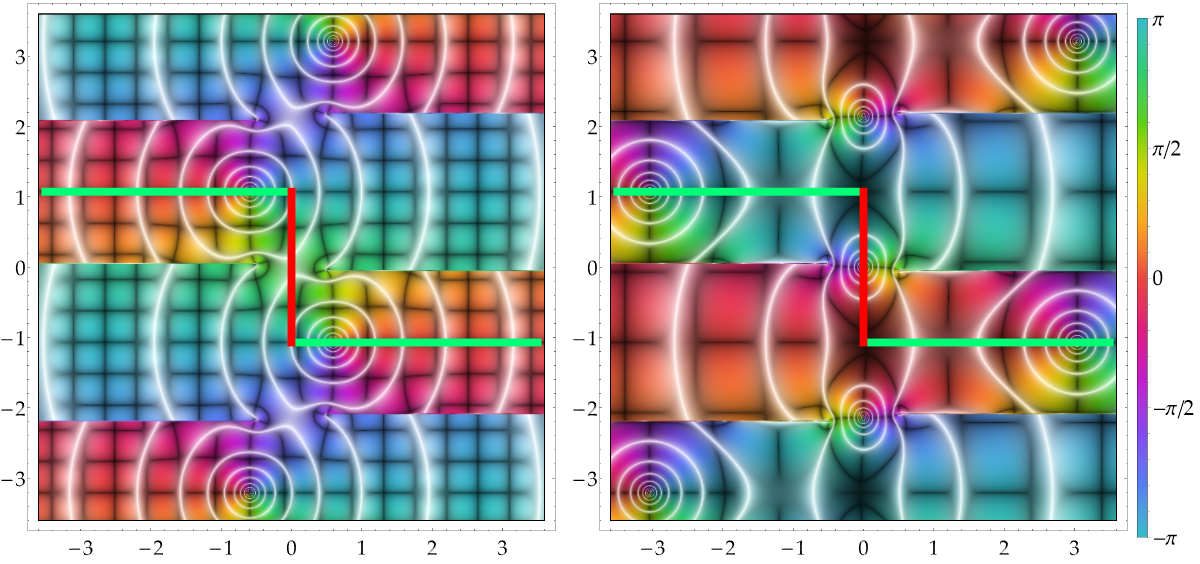}
    \caption{Complex instanton ($t(u)$ on the left, $z(u)$ on the right) for the field~\eqref{eq:field21} with parameters $\gamma_t = \omega_t/E = 1$ and $\gamma_z = \omega_z/E = 1$ for the first turning point $t(0) \approx -0.6064 + 0.8263 i$. The white lines are contour lines of $|q(u)|$, the black ones contour lines of the real/imaginary parts, and the color represents the phase according to the legend on the right. The red contour represents the creation region, where the instanton is complex, while the green contours are the acceleration regions, where the instanton is real. Evaluating the instantons in the acceleration regions and plotting them in the real $(t,z)$ plane yield Fig.~\ref{fig:Colliding}.}
    \label{fig:Complex}
\end{figure*}

As in~\cite{DegliEsposti:2023qqu}, after having found the turning point $t(0)$ we can solve the Lorentz-force equation along a dense set of proper-time contours $u(r)$ in the complex plane. While only a single contour is needed to obtain the spectrum, by considering this dense set we can see the analytical structure, e.g. where there are branch points or poles, which tells us what contour deformations are allowed, which is useful to know since, while equivalent contours give the same result, some contours can make the calculations simpler. For the field~\eqref{eq:field21}, which has two stationary points $(t,z) = (\pm \frac{1}{\sqrt{2} \omega}, 0)$, we find two independent instantons, one of which shown in Fig.~\ref{fig:Complex}.
\begin{figure}
    \includegraphics[width=\linewidth]{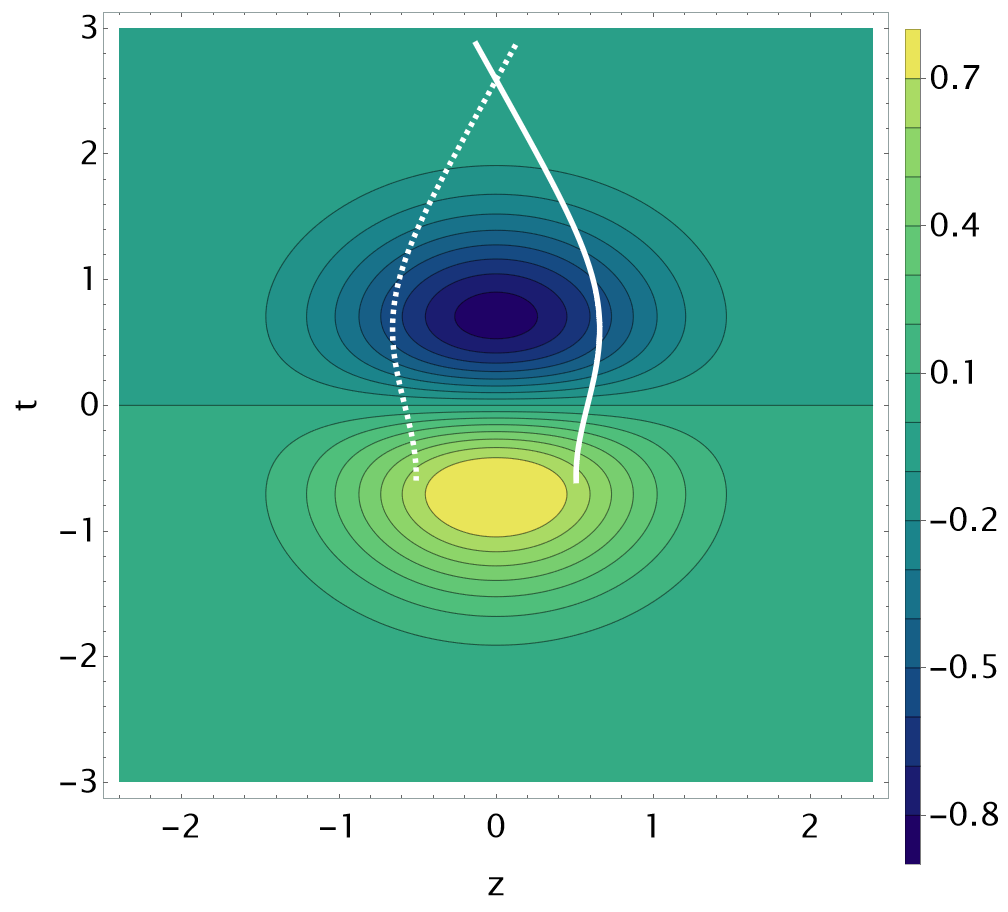}
    \caption{Contour plot shows the field~\eqref{eq:field21}. Instanton of Fig.~\ref{fig:Complex} evaluated in the acceleration regions (white lines) and background field~\eqref{eq:field21}. The continuous line represents the electron, the dashed line the positron. The pair, initially created and then repelled by the first pulse, is subsequently pulled together by the second pulse with opposite sign.}
    \label{fig:Colliding}
\end{figure}

We showed in~\cite{DegliEsposti:2023qqu} that, for $\Pi=\Pi_s$, there exists a special contour composed of a creation region, where the contour runs along the imaginary axis, and an acceleration region, which runs parallel to the real axis. While for the fields considered in~\cite{DegliEsposti:2023qqu} the $t$-component is imaginary and $z$ real in the creation region, in this case they are in general both complex. More importantly, however, both are purely real in the acceleration region. The contour is shown in Fig.~\ref{fig:Complex}, where the creation region is the vertical, red part of the contour, and the acceleration regions are the horizontal, green parts. Focusing on the acceleration regions, i.e. where the particles are real but are still inside the electromagnetic field, we get some intuition about the effect of the field by looking at the real worldline in the $(t,z)$ plane in Fig.~\ref{fig:Colliding}. The same pulse that creates the particles at $t<0$ also pushes them initially far apart. However, since the second pulse has opposite sign, the particles are pulled together and their worldlines cross.

\subsection{Instantons in the locally-constant-field limit}

For fields such as~\eqref{eq:field21}, we can find the turning point using a numerical continuation~\cite{Schneider:2018huk} starting from the time-dependent limit $\omega_z \to 0$. However, if we reverse the roles of time and space,
\be\label{eq:Field12}
E_3(t,z)/E_0 = -2\omega_z z \, e^{-(\omega_t t)^2 - (\omega_z z)^2} \;,
\ee
then the time-dependent limit is trivial and hence cannot be used as a starting point. On the other hand, taking the locally constant field limit gives a reasonable initial guess as well as a method of resolving the ambiguity in the sign of $z'(0) = \pm i$.  
We assume that the field has some stationary points $\{(t_n, z_n)\}$, e.g. $(\pm 1/\sqrt{2} \omega_t, 0)$ for the field~\eqref{eq:field21} or $(0,\sim\pm0.7/\omega_z)$ for~\eqref{eq:Field12}. Let $F_n = E_3(t_n, z_n)/E_0$ be the values of the field at the peaks. To take the $\omega\to0$ limit, we expand to leading order around a stationary point, which gives the Lorentz-force equation in a constant field, and find
\be
\begin{split}
    t(u) &= t_n + \frac{i}{|F_n|} \cosh(F_n u) \\
    z(u) &= z_n + \frac{i}{|F_n|} \sinh(F_n u) \;.
\end{split}
\ee
From this we obtain the constant-field turning point,
\be
t(0) = t_n +\frac{i}{|F_n|} \qquad z(0) = z_n \;,
\ee
which can then be used as an initial guess for the numerical continuation for non-constant fields, e.g.~\eqref{eq:Field12}. Furthermore, we see that the sign ambiguity in $z'(0)$ is solved, since
\be
z'(0) = i \; \mathrm{sign}(F_n)
\ee
in the constant-field limit, and if we increase the frequencies the sign will not suddenly flip. Thus, if we know which stationary point an instanton is continuously connected to, the sign of $z'(0)$ is given by the field evaluated at the peak.

\section{Integrated probability}

If we integrate~\eqref{eq:specTotal} with respect to all momenta (including transverse), we should obtain the same result as from the imaginary part of the effective action. When using the effective action for pair production, one is essentially working directly on the probability level and with closed worldlines~\cite{Dunne:2006st}.
Since our approach works at the amplitude level, for a field with $N$ stationary points we have $N$ instantons, compared to $N(N+1)/2$ instantons in the effective action method (due to the modulus squared of the $N$ amplitude terms and using symmetry~\eqref{eq:specTotal}). For example, if we have two peaks, in our approach we have two instantons on the amplitude level, but in the effective action approach one has one instanton for one peak, another instanton for the other peak, and then two additional instantons responsible for interference (complex conjugate of each other).

The closed instantons from the effective actions can be found as follows. As shown in~\cite{DegliEsposti:2023qqu}, the instantons are periodic along the imaginary proper-time $u$ axis at the momentum saddle point. So, while the contour shown by red and green lines in Fig.~\ref{fig:Complex} gives our open instantons, if we instead choose a contour along the imaginary axis then we recover the closed instantons~\cite{Dunne:2005sx,Dunne:2006st}  (this is another thing we can see clearly from plots such as Fig.~\ref{fig:Complex}). Similarly, for the $N$ independent open instantons at the saddle points of the diagonal terms of~\eqref{eq:specTotal} we find imaginary periodicity, as shown e.g. in Fig.~\ref{fig:Complex}.

On the other hand, as pointed out in~\cite{Dumlu:2011cc}, the contours associated to the closed instantons that are responsible for interference (i.e. the off-diagonal terms of~\eqref{eq:specTotal}) are nontrivial. For such terms the exponent
\be\label{eq:ComplexPsi0}
\psi_a(\Pi) + [\psi_b(\Pi)]^*
\ee
is in general complex even for $\Pi \in \mathbb R$, and the momentum saddle point will in general also be complex. However, \eqref{eq:ComplexPsi0} is not analytic for complex $\Pi$ due to the complex conjugation. Instead we should use
\be\label{eq:ComplexPsi}
\Psi_{ab}(\Pi) := \psi_a(\Pi) + [\psi_b(\Pi^*)]^*
\ee
as an analytic continuation of $\psi_a(\Pi) + \psi_b(\Pi)^*$ for $\Pi \in \mathbb C$.

To find the (complex) saddle point of $\Psi(\Pi)$ and the instantons, let us focus on symmetric fields for simplicity; for generic fields one simply doubles the conditions and the free parameters by considering conditions at $u = -\infty$ in addition to conditions at $u=+\infty$, and $z(0)$ in addition to $t(0)$. Since the exponents $\psi_a$ and $\psi_b$ in~\eqref{eq:ComplexPsi} are evaluated at momenta that are the complex conjugate of each other, we must have
\be
\Re \, z_a'(\infty) = \Re \, z_b'(\infty) \qquad \Im \, z_a'(\infty) = -\Im \, z_b'(\infty) \; .
\ee
Setting both the real and the imaginary parts of the derivative of~\eqref{eq:ComplexPsi} to zero, the conditions to find the instantons at the complex momentum saddle point are
\be\label{eq:ComplexP}
\begin{split}
    &\Re \, z_a'(\infty) = \Re \, z_b'(\infty) \qquad \Im \, z_a'(\infty) = -\Im \, z_b'(\infty) \\
    &\Im \left[ \frac{z_a'(\infty)}{t_a'(\infty)}t_a(\infty) - z_a(\infty) -\frac{z_b'(\infty)}{t_b'(\infty)}t_b(\infty) + z_b(\infty) \right] = 0 \\
    &\Re \left[ \frac{z_a'(\infty)}{t_a'(\infty)}t_a(\infty) - z_a(\infty) + \frac{z_b'(\infty)}{t_b'(\infty)}t_b(\infty) - z_b(\infty)\right] = 0
\end{split}
\ee
with two complex free parameters $t_a(0)$ and $t_b(0)$. The different sign between the real/imaginary parts is due to the complex conjugation in $[\psi_b(\Pi^*)]^*$.

\begin{figure*}
    \includegraphics[width=\linewidth]{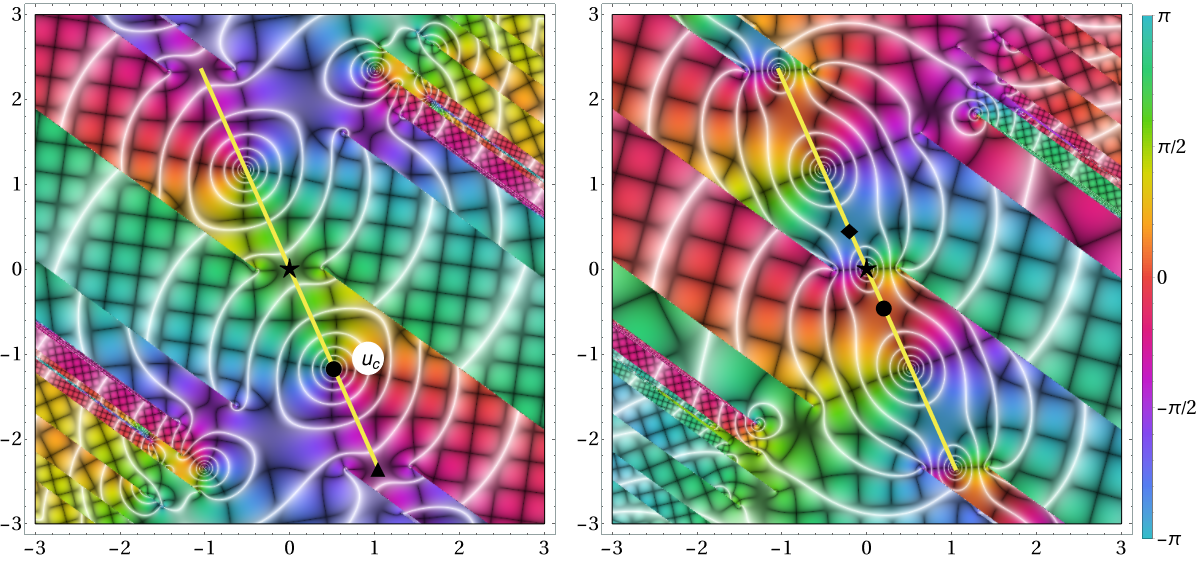}
    \caption{One of the two complex-momentum instantons ($t(u)$ on the left, $z(u)$ on the right) for the field~\eqref{eq:field21} with parameters $\gamma_t = \omega_t/E_0 = 1$ and $\gamma_z = \omega_z/E_0 = 1$. The complex momenta for these parameters are $p_3 \approx 0.9994 + 0.2678 i$ and $p'_3 = -p_3$. Along the yellow line the instanton components are periodic. In this particular example, the space $z$-component is doubly periodic along the yellow contour.}
    \label{fig:qIntComplex}
\end{figure*}

\begin{figure*}
    \raisebox{.1cm}{\includegraphics[height=5.2cm]{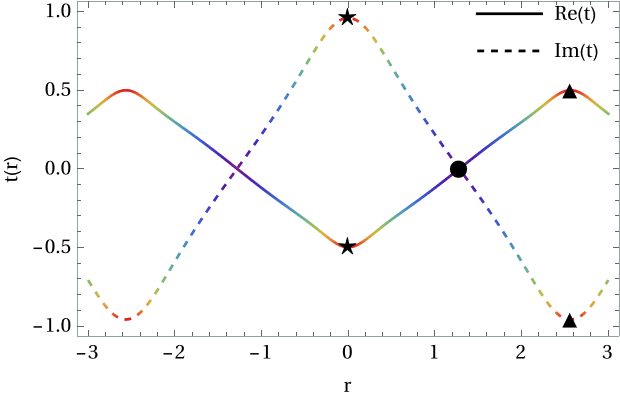}}
    \hfill
    \includegraphics[height=5.35cm]{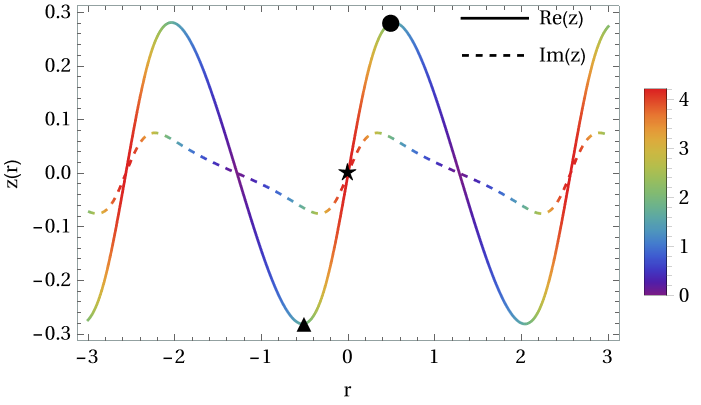}
    \caption{Instanton components evaluated along the yellow contour in Fig.~\ref{fig:qIntComplex} ($u(r)\in\mathbb{C}$ parametrized by $r\in\mathbb{R}$). The black markers are related to the ones on the complex plane in~\ref{fig:qIntComplex}. The plots are colored according to the value of the field $|E(t(r),z(r))|/E_0$.}
    \label{fig:Tilted}
\end{figure*}

We can see some instantons obtained with this method\footnote{These instantons always come in pairs; in the picture we see one of the two, and the other is very similar.} on the complex plane in Fig.~\ref{fig:qIntComplex}. The periodic contour is shown in yellow, and the instantons evaluated on such contour are shown in Fig.~\ref{fig:Tilted}.

\section{Interference from separated identical pulses}

We have described above how to obtain the spectrum for quite general fields with multiple peaks. In this section we will consider sets of identical peaks which are well separated, which will allow us to gain further insights into the interference patterns shown above, and use them as a guide for how to design field configurations to achieve some desired spectrum pattern.

Consider first a field given by two identical pulses,
\be\label{FsumF1}
F(t,z)=F_1(t-t_a,z-z_a)+F_1(t-t_b,z-z_b) \;,
\ee
where
\be
F_1(t,z)=F_z(\gamma_z z)F_t(\gamma_t t)
\ee
and e.g. $F_z(v)=e^{-v^2}$, $\text{sech}^2(v)$ or some asymmetric pulse peaked such as 
\be
F_z(v)=\frac{2e^{-v^2}}{1+e^{-v^3}} \;,
\ee
which still only has a single peak but which is much steeper on one side.
Similarly for $F_t$. We assume that $\Delta t=t_b-t_a$ and/or $\Delta z=z_b-z_a$ are sufficiently large that there is negligible overlap between the two pulses. Then the instanton created around pulse $b$, $q^{(b)}_\mu(u)$, is essentially the same as the instanton created around pulse $a$, $q^{(a)}_\mu$. In particular, $z_{(a)}'(\infty)=z_{(b)}'(\infty)$, so both instantons give amplitudes centered around the same asymptotic momentum. The only difference is a constant shift, $q^{(a)}_\mu(u)=q_\mu(u)+q^a_\mu$ and $q^{(b)}_\mu(u)=q_\mu(u)+q^b_\mu$, where $q_\mu(u)$ is the instanton for $F_1(t,z)$, and $q^a_\mu$ and $q^b_\mu$ are constant. 
We are implicitly assuming that the instanton created at one pulse does not pass through the other pulse. It is of course possible for pulse $b$ to be in the asymptotic path of the instanton created at pulse $a$, and then $q_{(b)}(u)-q_{(a)}(u)\ne q_b-q_a$, but this requires specific values of $\Delta z$ and $\Delta t$. In the generic case, the instanton from one pulse misses the other pulse. See more below.  

In contrast, if we have a purely time dependent field and $t_a<t_b$, then the instanton created by pulse $a$ will always go through pulse $b$ regardless of how far apart the pulses are. It was shown in~\cite{Akkermans:2011yn} using a WKB approach that in this case there is no interference if the two pulses have the same sign, but there is interference if they have opposite sign, which is the complete reverse of what we have for a spacetime field. In the instanton picture we can understand this as follows. If the two purely time-dependent pulses have the same sign, the electron or positron half of the $q_{(a)}$ instanton will accelerate through about half of pulse $a$ and all of pulse $b$, so the two instantons have different asymptotic momenta ($z_{(a)}'(\infty)\sim 3z_{(b)}'(\infty)$), which means the instantons will not lead to interference. However, if the two pulses have opposite sign, $F(q_a)>0$ and $F(q_b)<0$, then $z_{(a)}'$ will be reflected by the second pulse so that $z_{(a)}'(\infty)=z_{(b)}'(\infty)$ and interference does happen.       
Thus, including the $z$ dependence gives a fundamentally different case compared to purely time-dependent fields considered in e.g.~\cite{Akkermans:2011yn}. Previous studies using the Wigner approach, \cite{Kohlfurst:2015zxi,Ababekri:2019dkl} noticed that interference patterns in the time-dependent case disappear if the field is sufficiently short in the spatial direction.

In general, when there are more than one term in the amplitude, one needs to compute the eigenvalues to fix the Maslov-Morse index (cf.~\ref{sec:eigenvalues}). However, from the eigenvalue equation~\eqref{eq:eigenEq} one sees that the eigenvalues are the same for the two peaks, because the instantons are equal up to a constant shift, so then the index is the same for both amplitude terms and therefore just gives an overall phase which drops out upon squaring the amplitude.

\subsection{Moir\'e patterns}

Denoting $\Delta\alpha=\alpha^b-\alpha^a$ and similarly for other quantities, we find from~\eqref{eq:alphas}
\be\label{alphaPbma}
\Delta\alpha_P=\Delta t\left(\frac{p'_3}{p'_0}-\frac{p_3}{p_0}\right)
\ee
and
\be\label{alphaDeltabma}
\Delta\alpha_\Delta=\Delta z+\frac{\Delta t}{2}\left(\frac{p'_3}{p'_0}+\frac{p_3}{p_0}\right) \;.
\ee    
The $\delta q$ solutions are the same for both pulses, so from~\eqref{d2andbetafromX} we find that the widths $d_{ij}^{-2}$ are also the same for both pulses, while 
\be\label{betaDeltaDeltabma}
\Delta\beta_{\Delta\Delta}=\frac{\Delta t}{4}\left(\frac{1}{p_0^{\prime3}}+\frac{1}{p_0^3}\right)
\ee
\be\label{betaPPbma}
\Delta\beta_{PP}=\Delta t\left(\frac{1}{p_0^{\prime3}}+\frac{1}{p_0^3}\right)
\ee
and
\be\label{betaPDeltabma}
\Delta\beta_{P\Delta}=\frac{\Delta t}{2}\left(\frac{1}{p_0^{\prime3}}-\frac{1}{p_0^3}\right) \;.
\ee

\begin{figure*}
    \centering
    \includegraphics[width=.45\linewidth]{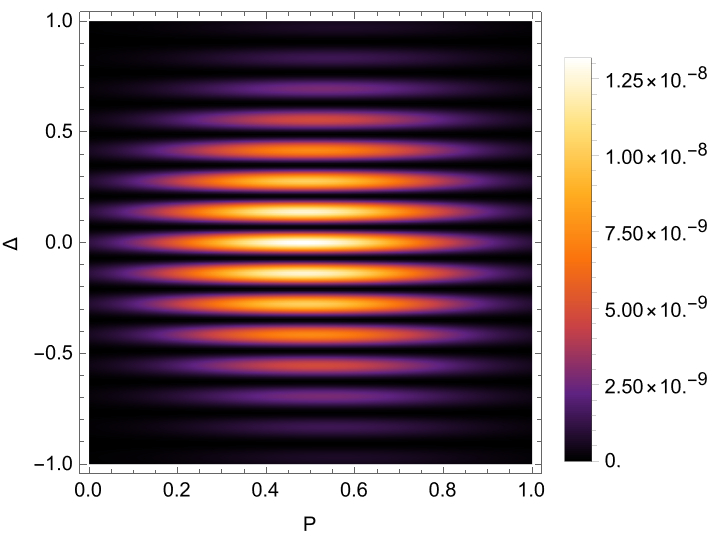}
    \includegraphics[width=.45\linewidth]{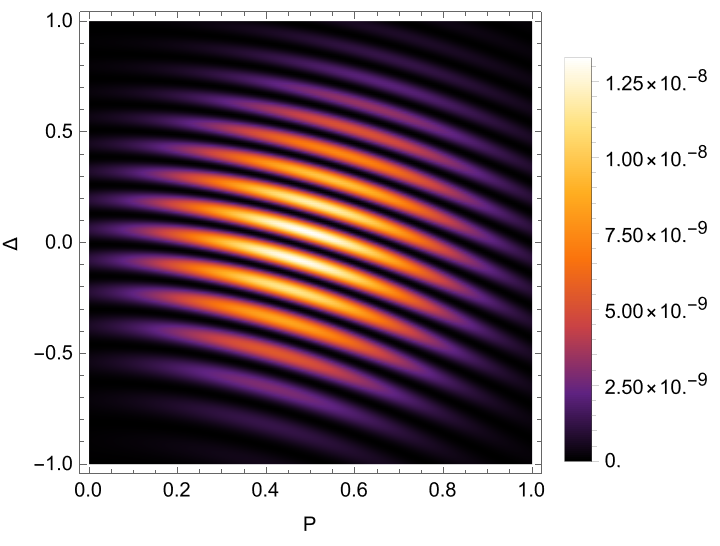}\\
    \includegraphics[width=.45\linewidth]{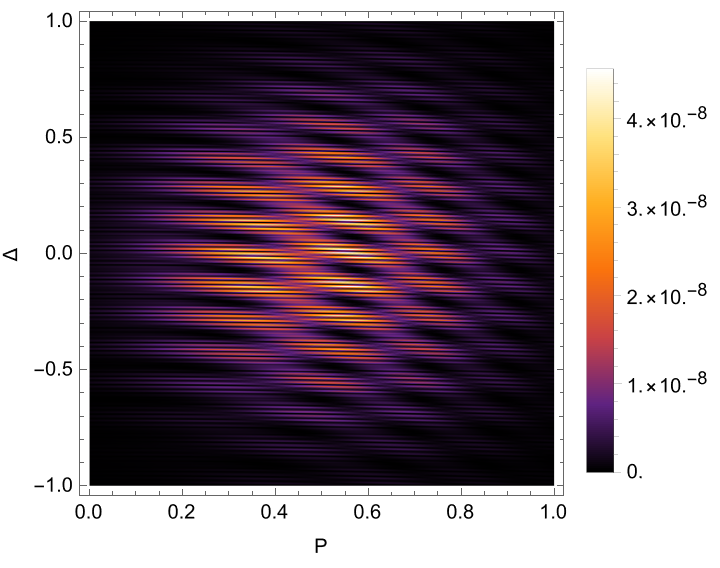}
    \includegraphics[width=.45\linewidth]{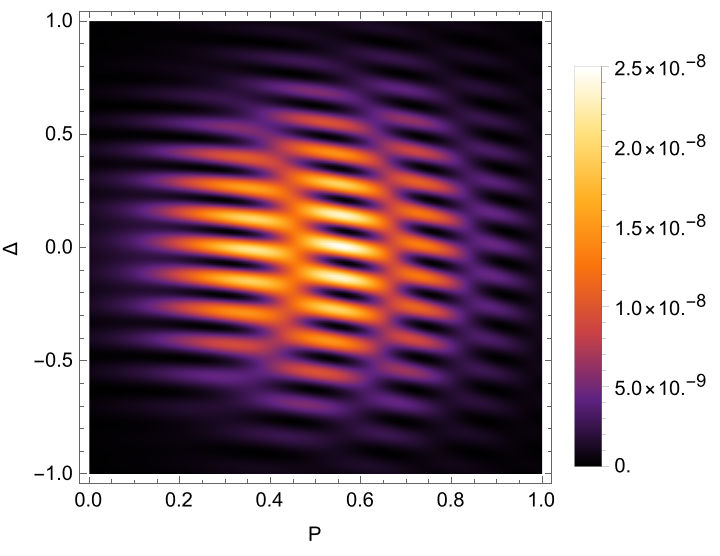}
\caption{Moir\'e patterns in the spectrum produced by four Gaussian peaks~\eqref{fourPulses} placed as in~\eqref{fourPulsesPositions}, with $\gamma_t=\gamma_z=1$ and $E_0=0.2$. The first (second) plot is obtained with only the $a$ and $b$ ($c$ and $d$) pulses. The third (fourth) plot is obtained by adding the contribution from all four pulses using $|e^{\psi_a}+e^{\psi_b}+e^{\psi_c}+e^{\psi_d}|^2$ ($|e^{\psi_a}+e^{\psi_b}|^2+|e^{\psi_c}+e^{\psi_d}|^2$) The results have been obtained using the grid method, but, as shown in Fig.~\ref{fig:moire4z1D}, the quadratic approximation~\eqref{eq:expExpanded} gives basically the same results.}
\label{fig:moire4z}
\end{figure*}

Since the oscillations are mostly due to ${\bm\alpha}$ rather than ${\bm\beta}$, the two amplitude terms, $a$ and $b$, will lead to more or less parallel stripes as interference patterns in the $P$-$\Delta$ plane, where the stripes are orthogonal to ${\bm\alpha}$. Since only $\Delta\alpha_\Delta$ depends on $\Delta z$, we find that we have the freedom to choose any angle of ${\bm\alpha}$ by increasing or decreasing $\Delta t$ and $\Delta z$. If we choose $\Delta t\ne0$ and $\Delta z=0$ then we find stripes parallel to the $\Delta$ axis, and if we instead choose $\Delta t=0$ and $\Delta z\ne0$ then we find stripes parallel to the $P$ axis. If we have more than two terms, we can have interference between stripes along one angle and other stripes along some other angle, which gives moir\'e patterns, e.g. as in Fig.~\ref{fig:moire4z}.  

\begin{figure*}
    \centering
    \includegraphics[width=.8\linewidth]{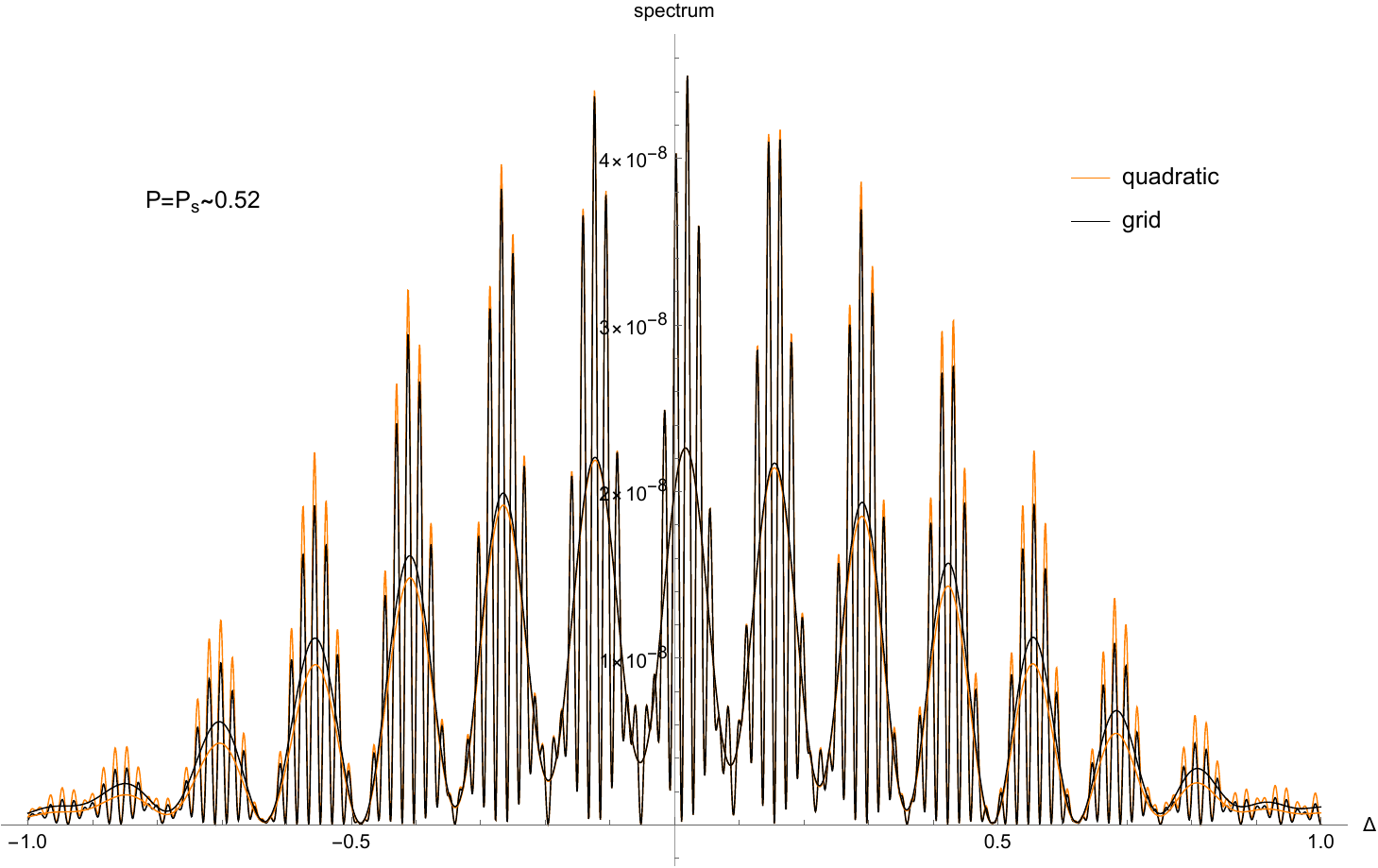}
    \caption{The ``grid'' lines are obtained by computing instantons at each point on a $10\times10$ grid over the rectangle $0<P<1$ and $-1<\Delta<1$. The ``quadratic'' lines are obtained using the quadratic approximation~\eqref{eq:expExpanded} around the saddle point, $P_s\approx0.52$ and $\Delta_s=0$. The lines with faster oscillations are the full result $|e^{\psi_a}+e^{\psi_b}+e^{\psi_c}+e^{\psi_d}|^2$, while the lines which look like a moving average are given by $|e^{\psi_a}+e^{\psi_b}|^2+|e^{\psi_c}+e^{\psi_d}|^2$.}
    \label{fig:moire4z1D}
\end{figure*}

\subsection{Aharonov-Bohm phase}

The following contribution is more nontrivial, 
\be
-i\psi=\int\ud u\,q^\mu\partial_\mu A_\nu\frac{\ud q^\nu}{\ud u} \;.
\ee
Using the Lorentz-force equation and partial integration\footnote{The boundary terms $q^\mu A_\mu|_{-\infty}^\infty$ vanishes for spacetime fields.} gives 
\be\label{tildePhi1}
-i\psi=\int\ud u(qq''-q'_\mu A^\mu) \;.
\ee
So for
\be
A_\mu(q)=A^{(1)}_\mu(q-q_a)+A^{(1)}_\mu(q-q_b)
\ee
we find
\be
-i\Delta\psi=\Delta t(p_0+p'_0)+\Delta z(p_3+p'_3)+\psi_{\rm AB} \;,
\ee
where 
\be
\psi_{\rm AB}=-\int\ud u\, q'\Delta A \;,
\ee
$\Delta q^\mu=q_b^\mu-q_a^\mu$ and
\be
\Delta A_\mu=A^{(1)}_\mu(q+\Delta q)-A^{(1)}_\mu(q-\Delta q) \;.
\ee
$\psi_{\rm AB}$ is an Aharonov-Bohm term~\cite{Aharonov:1959fk,EhrenbergSiday}. Using Stokes' theorem we can rewrite it as
\be
\psi_{\rm AB}=-\int_R\ud t\ud z\, F(t,z) \;,
\ee
where $R$ is the region between the curves $q(u)+\Delta q$ and $q(u)-\Delta q$. Since we have assumed that $\Delta q$ is sufficiently large so that the two pulses do not overlap, the precise shape of the boundary of $R$ is irrelevant; either $R$ covers the region where $F$ is nonzero, and then we have
\be
\psi_{\rm AB}=-\text{sign}(\Delta t)\int_{-\infty}^\infty\ud t\ud z\, F(t,z) \;,
\ee
or otherwise $\psi_{\rm AB}=0$. 
Since $t(\pm\infty)=\infty$ while $z(\pm\infty)=\pm\infty$, $\psi_{\rm AB}$ is only nonzero if $\Delta t\ne0$. Thus, if we have two pulses that are separated by a nonzero $\Delta z$ but $\Delta t=0$, then $\psi_{\rm AB}=0$. 

Note that $\Delta\psi$ is purely imaginary, so $\Delta\mathcal{A}=0$ and $\Delta\phi=-i\Delta\psi$. In other words, while the two amplitude terms have different phases, the real part of the exponent, which gives the order of magnitude of the amplitude term, is the same.

\subsection{Fabry-Perot patterns}

Thus, the term in the probability amplitude coming from pulse $a$ differs by a pure phase compared to the term from pulse $b$, and the probability is given by
\be
\begin{split}
&\mathbb{P}(\text{both pulses})=2\mathbb{P}(\text{one pulse})\bigg\{1+\cos\bigg(\Delta\phi\\
&+\Delta{\bm\alpha}\cdot[{\bm\Pi}-{\bm\Pi}_s]
+\frac{1}{2}[{\bm\Pi}-{\bm\Pi}_s]\cdot\Delta{\bm\beta}\cdot[{\bm\Pi}-{\bm\Pi}_s]\bigg)\bigg\} \;.
\end{split}
\ee
$\Delta{\bm\alpha}$ and $\Delta{\bm\beta}$ only depend on the separation of the two pulses, $\Delta q$, and the asymptotic momenta of the instantons, while $\Delta\phi$ also has the Aharonov-Bohm term. 

It is straightforward to generalize the above result to $N$ equal pulses which are equally separated using
\be\label{FabryPerotEqual}
\left|\sum_{n=0}^{N-1}e^{2i\theta n}\right|^2=\frac{\sin^2(N\theta)}{\sin^2\theta} \;.
\ee
Such Fabry-Perot interference patterns were studied for time-dependent fields in~\cite{Akkermans:2011yn} (and in~\cite{Ilderton:2019ceq} for nonlinear Breit-Wheeler pair production by plane waves). However, what we have is quite different from the time-dependent case. If we consider a train of pulses with alternating signs, $E>0$ or $E<0$, then there is no interference between two pulses of different signs because they lead to different $P_s$ and $\Delta_s$, so if say $P$ is close to $P_s(E>0)$ then it is far from $P_s(E<0)$ and hence the overlap is exponentially small. Thus, in the region of the spectrum close to $P_s(E>0)$ we only need to consider the $E>0$ pulses, and vice versa for the region around $P_s(E<0)$.

\subsection{Moir\'e patterns}

As an example of moir\'e patterns, we consider four separated Gaussian peaks
\be\label{fourPulses}
F=F_1(q-q_a)+F_1(q-q_b)+F_1(q-q_c)+F_1(q-q_d) \;,
\ee
where $F_1=\exp[-(\omega_t t)^2-(\omega_z z)^2]$. We focus on $\omega_t=\omega_z$. We place the peaks at
\be\label{fourPulsesPositions}
\begin{split}
(t_a,z_a)&=(0,0) \quad \gamma(t_b,z_b)=(0,9) \\
\gamma(t_c,z_c)&=(0,70) \quad \gamma(t_d,z_d)=(6,79) \;.
\end{split}
\ee
From~\eqref{alphaPbma} and~\eqref{alphaDeltabma}, we expect that, since $\Delta t=0$ and $\Delta z\ne0$ for the $a$ and $b$ peaks, they alone will produce oscillations in the $\Delta$ direction, which is also what we see in the first plot in Fig.~\ref{fig:moire4z}. For the $c$ and $d$ peaks we have both $\Delta t\ne0$ and $\Delta z\ne0$, so they alone produce oscillations at an angle from the $\Delta$ direction, as in the second plot in Fig.~\ref{fig:moire4z}. If we just add the pattern from the $c$ and $d$ peaks onto the pattern from the $a$ and $b$ peaks then we would expect to see moir\'e patterns. The spectrum has in general a more complicated shape because
\be
|e^{\psi_a}+e^{\psi_b}+e^{\psi_c}+e^{\psi_d}|^2\ne
|e^{\psi_a}+e^{\psi_b}|^2+|e^{\psi_c}+e^{\psi_d}|^2 \;.
\ee
However, since we have chosen the $c$ and $d$ pulses to be very far from the $a$ and $b$ pulses, the cross terms, $2\cos(\text{Im}[\psi_a-\psi_c])$ etc., oscillate very rapidly, as can be seen in Fig.~\ref{fig:moire4z1D}. If the oscillations are faster than the detector's resolution, then we can neglect the cross terms, $|(a)+(b)+(c)+(d)|^2\to |(a)+(b)|^2+|(c)+(d)|^2$, which basically gives a moving average. The resulting moir\'e patterns, with and without the cross terms, are shown in Fig.~\ref{fig:moire4z}. 

\begin{figure}
    \centering
    \includegraphics[width=\linewidth]{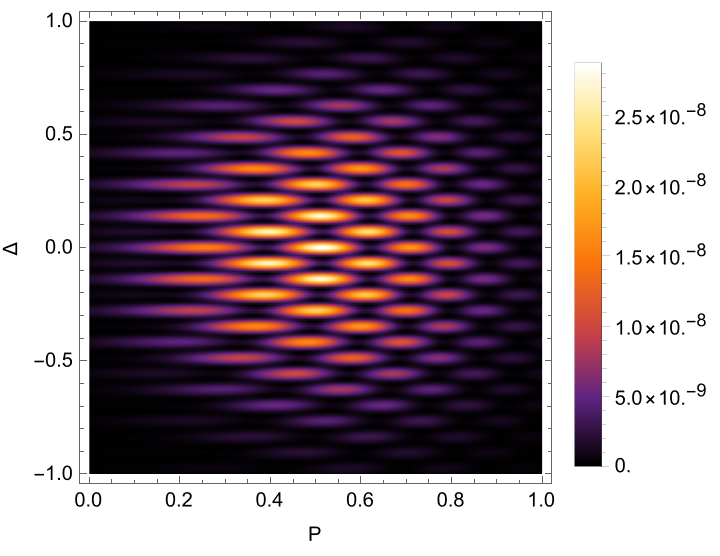}
    \caption{Moir\'e pattern in momentum spectrum for three separated Gaussian peaks~\eqref{threePulses} placed at~\eqref{threePulsesPositions}, for $\gamma_t=\gamma_z=1$ and $E=0.2$. The results have been obtained with the grid approach, but as Fig.~\ref{fig:moire3PandDelta} shows, the quadratic approximation~\eqref{eq:expExpanded} gives basically the same result.}
    \label{fig:moire3z2D}
\end{figure}

\begin{figure}
    \centering
    \includegraphics[width=\linewidth]{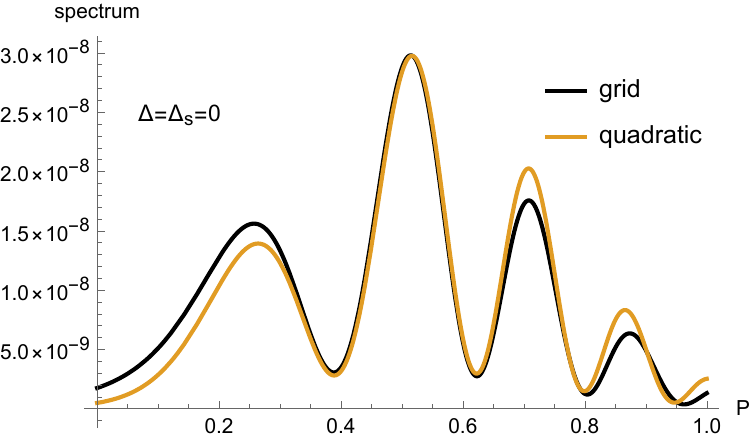}
    \includegraphics[width=\linewidth]{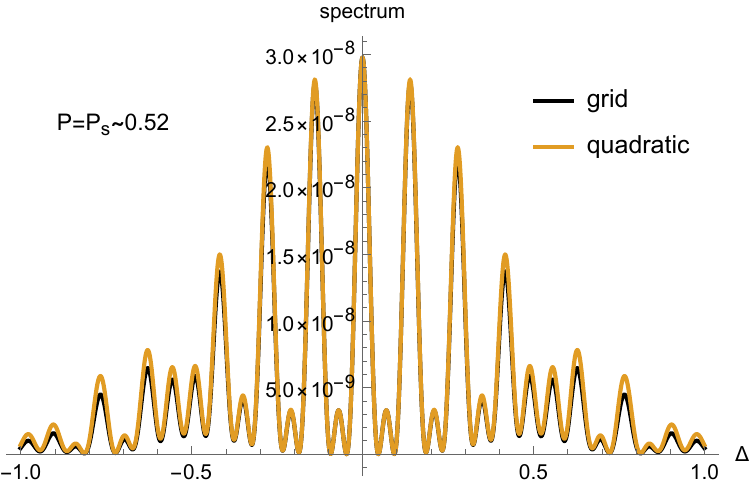}
    \caption{Same case as in Fig.~\ref{fig:moire3z2D}. The ``grid'' lines are obtained by computing instantons at each point on a $10\times10$ grid over the rectangle $0<P<1$ and $-1<\Delta<1$. The ``quadratic'' lines are obtained using the quadratic expansion~\eqref{eq:expExpanded} around the saddle point, $P_s\approx0.52$ and $\Delta_s=0$.}
    \label{fig:moire3PandDelta}
\end{figure}

We chose four pulses in the above example in order to have two pairs and hence two patterns that can form a combined pattern in a way similar to what one might usually think of as moir\'e patterns. We can though obtain 2D patterns in the $p_3-p'_3$ plane with only three pulses too. As an example we consider
\be\label{threePulses}
F=F_1(q-q_a)+F_1(q-q_b)+F_1(q-q_c) \;,
\ee
with the same Gaussian $F_1$ as before. We choose
\be\label{threePulsesPositions}
(t_a,z_a)=(0,0) \quad \gamma(t_b,z_b)=(6,9) \quad \gamma(t_c,z_c)=(6,-9) \;.
\ee
Fig.~\ref{fig:moire3PandDelta} compares the results for the spectrum obtained using either the grid approach or the quadratic expansion~\eqref{eq:expExpanded} around the saddle point. The agreement is good, especially given that the plot stretches over a region where $\Pi-\Pi_s=\mathcal{O}(1)$ is not actually small.

\subsection{Notes on the grid approach}

\begin{figure}
    \centering
    \includegraphics[width=\linewidth]{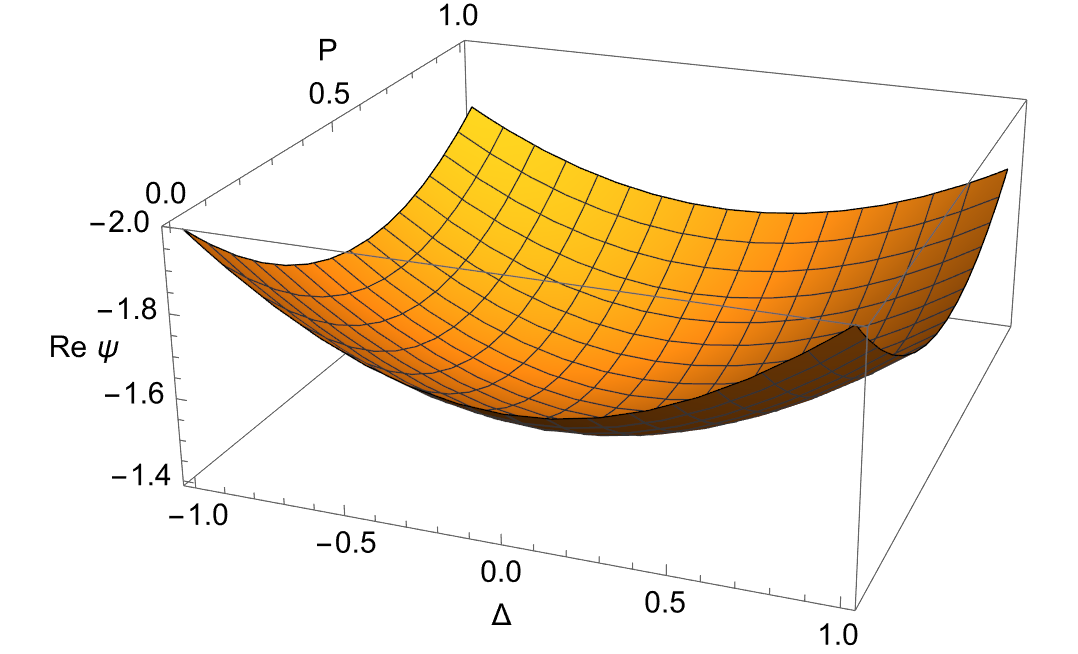}\\
    \includegraphics[width=\linewidth]{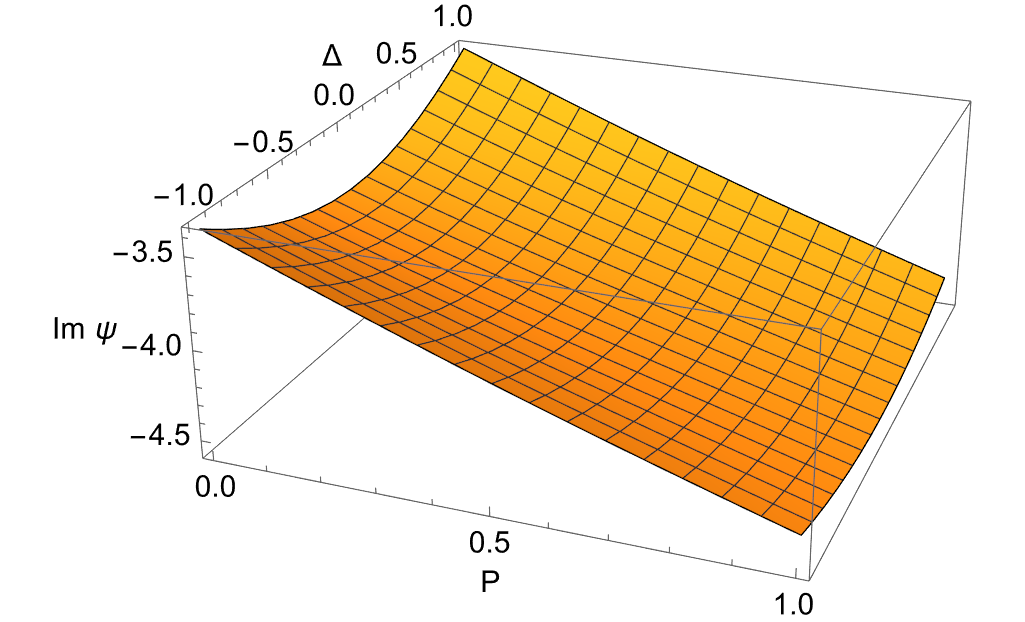}\\
    \includegraphics[width=\linewidth]{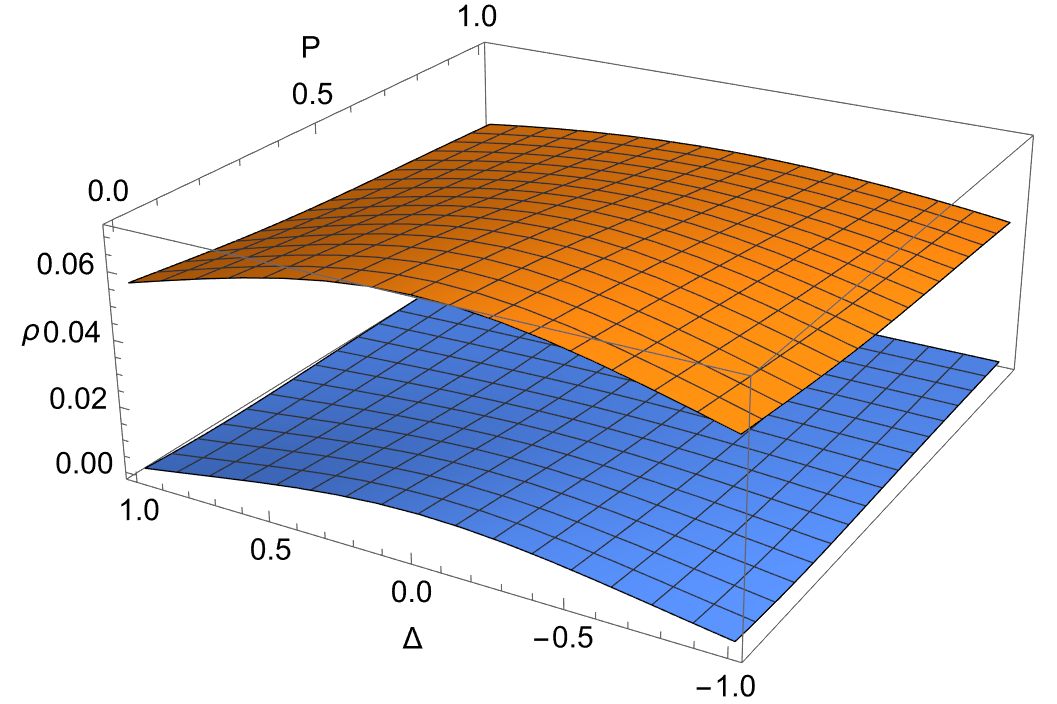}
    \caption{Ingredients for pulse $a$ for the example in Fig.~\ref{fig:moire3z2D}. The third plot shows the real (orange/upper) and imaginary (blue/lower) parts of the pre-exponential factor, $\rho=\sqrt{\frac{4\pi}{(2\pi)^4p_0p_0' h}}$.}
    \label{fig:actuallySlow}
\end{figure}

Just looking at the plots above for the spectrum, one sees rapid or at least several oscillations, and might therefore be led to think that one would need a large number of points in $\Pi$ for the grid approach to capture all the oscillations. However, we do not actually need that many points. Fig.~\ref{fig:actuallySlow} shows the ingredients for pulse $a$ for the example in Fig.~\ref{fig:moire3z2D}. These functions are slowly varying without oscillations, so we only need to evaluate them at relatively few points in order to make a precise interpolation function. For the particular example in Fig.~\ref{fig:actuallySlow} we have used a grid with $10\times10$ points in $0<P<1$ and $-1<\Delta<1$, and third-order polynomial interpolation (using Mathematica's built-in ``Interpolation'' function). 

Even though we do not need a large number of grid points, evaluating the integral in~\eqref{eq:psiIntegral} at each point could still be somewhat time consuming, or at least unnecessarily time consuming. We can compute $\psi$ much faster using the fact that the gradient, $\nabla\psi=(\partial_P,\partial_\Delta)\psi$, can be obtained without doing any integrals (cf.~\eqref{eq:alphas})
\be
\begin{split}
\partial_P\psi&=i\left[-z(u_1)+z(u_0)-\frac{p_3}{p_0}t(u_1)+\frac{p_3'}{p_0'}t(u_0)\right]\\
\partial_\Delta\psi&=\frac{i}{2}\left[z(u_1)+z(u_0)+\frac{p_3}{p_0}t(u_1)+\frac{p_3'}{p_0'}t(u_0)\right] \;.
\end{split}
\ee
We evaluate $\nabla\psi$ on each grid point and make an interpolation function of the table, from which we can then obtain $\psi$ using
\be\label{psiFromGradPsi}
\psi(\Pi)=\psi(\Pi_s)+\int_{\Pi_s}^\Pi\ud\tilde{\Pi}\cdot\nabla\psi(\tilde{\Pi}) \;,
\ee
where the line integral can be performed e.g. along a straight line
\be
\tilde{\Pi}(s)=(1-s)\Pi_s+s\Pi \;.
\ee
Now we only have to perform the integral in~\eqref{eq:psiIntegral} once, to obtain $\psi(\Pi_s)$. For the other grid points we just have to perform the integral in~\eqref{psiFromGradPsi}, which is very fast. This gives a table of $\psi$ values on the grid. Making an interpolation function of that table gives the final result.

\subsection{LCF limit}

For $\gamma\ll1$ we showed in~\cite{DegliEsposti:2023qqu} that the final momentum can be obtained by using the fact that the instanton travels at close to the speed of light during most of the acceleration region. This allows us to find a simple approximation for the asymptotic momentum without having to find any instanton. For the positron we have
\be
p'_0=-t'(-\infty)\approx p'_3=z'(-\infty)\approx\int_{-\infty}^0\!\ud t\, F_z(\gamma_z t)F_t(\gamma_t t) \;,
\ee
and for the electron
\be
p_0=t'(\infty)\approx-p_3=z'(\infty)\approx\int_0^\infty\ud t\, F_z(\gamma_z t)F_t(\gamma_t t) \;.
\ee
So $p_0,p'_0=\mathcal{O}(1/\gamma)$ and the pair has high energy. Inserting this into~\eqref{alphaPbma} gives
\be
\Delta\alpha_P\approx2\Delta t=\mathcal{O}(1/\gamma) \;,
\ee
where we have kept $\gamma_t\Delta t$ fixed so that the two pulses stay separated. Thus, the frequency of the interference pattern increases as $1/\gamma$ in the LCF limit. From~\eqref{betaDeltaDeltabma}, \eqref{betaPDeltabma} and~\eqref{betaPDeltabma} we see that $\Delta\beta=\mathcal{O}(\gamma^2)$. However, we also have $d_{ij}^{-2}=\mathcal{O}(\gamma^2)$, so the natural scale for a plot of the spectrum is $\Pi-\Pi_s=\mathcal{O}(1/\gamma)$, so $(\Pi-\Pi_s)\cdot\Delta{\bm\beta}\cdot(\Pi-\Pi_s)=\mathcal{O}(1)$. Although this is much smaller than $\Delta{\bm\alpha}\cdot(\Pi-\Pi_s)=\mathcal{O}(1/\gamma^2)$, a $\mathcal{O}(1)$ term cannot be dropped when it appears in $\cos(\dots)$. However, since $\Delta{\bm\alpha}\cdot(\Pi-\Pi_s)=\mathcal{O}(1/\gamma^2)$ the oscillations are very rapid for $\gamma\ll1$ and so, given a finite resolution of the detector, eventually the entire $\cos(\dots)$ term can be neglected since it averages out. Similarly, for a train of $N$ equal and equally separated pulses, the oscillating terms in~\eqref{FabryPerotEqual} average out and we are left with $\mathbb{P}(N\text{ pulses})\approx N\mathbb{P}(\text{one pulse})$. Thus, the probability becomes effectively incoherent in the LCF limit.  

\subsection{Scattering}

\begin{figure*}[h]
    \includegraphics[width=0.49\linewidth]{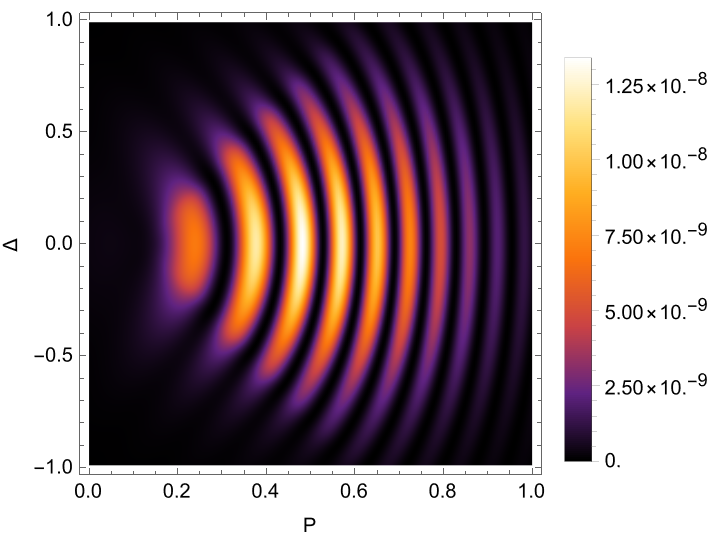}
    \includegraphics[width=0.49\linewidth]{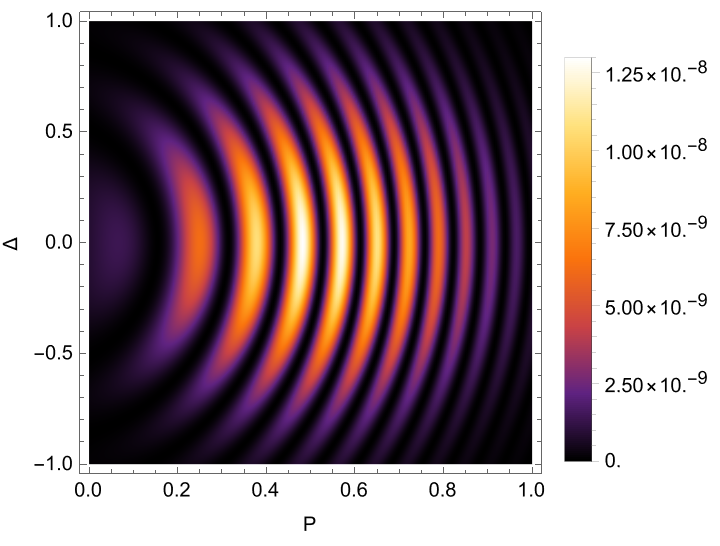}\\
    \includegraphics[width=0.49\linewidth]{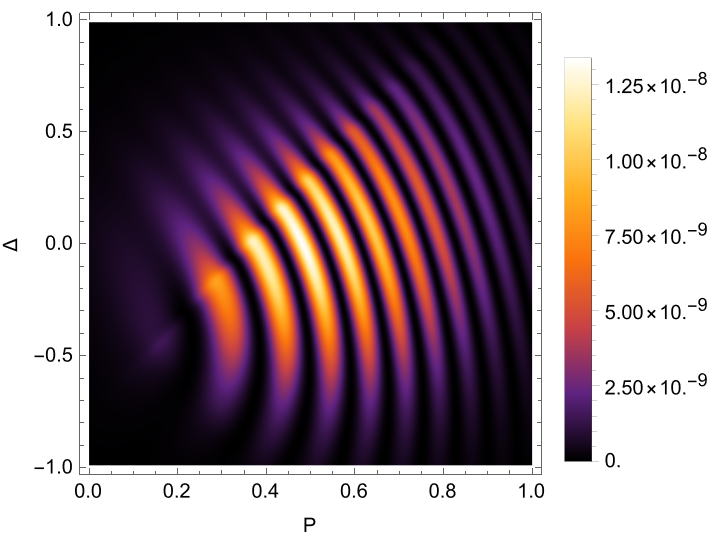}
    \includegraphics[width=0.49\linewidth]{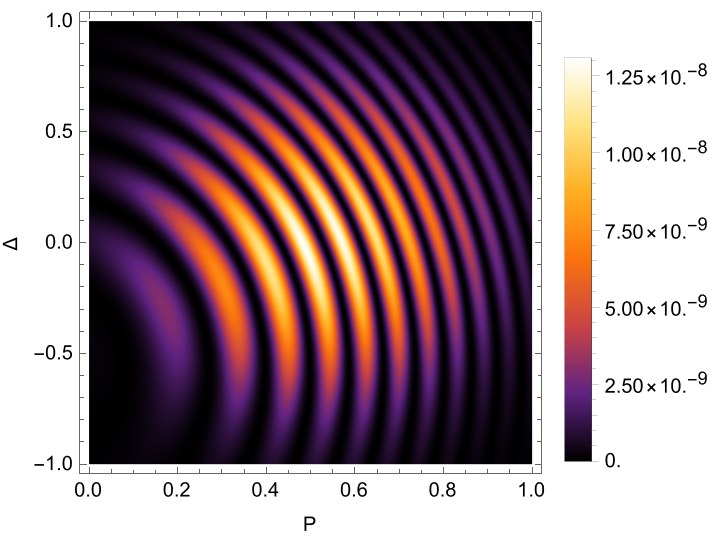}\\
    \includegraphics[width=0.49\linewidth]{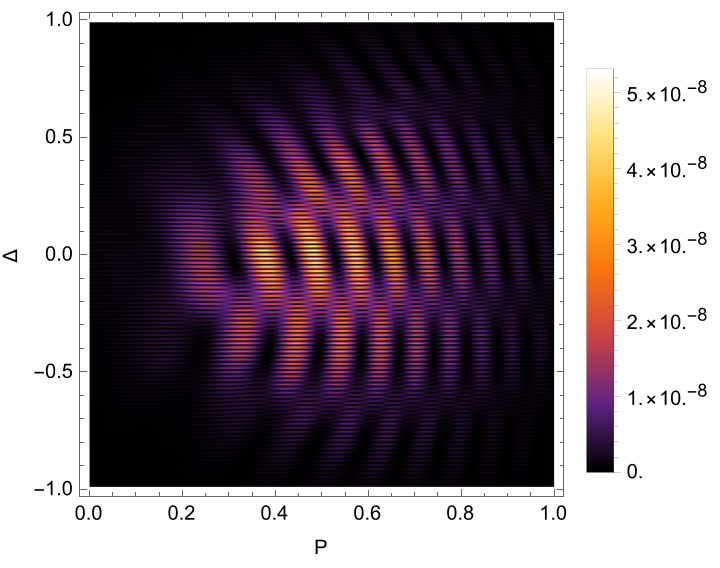}
    \includegraphics[width=0.49\linewidth]{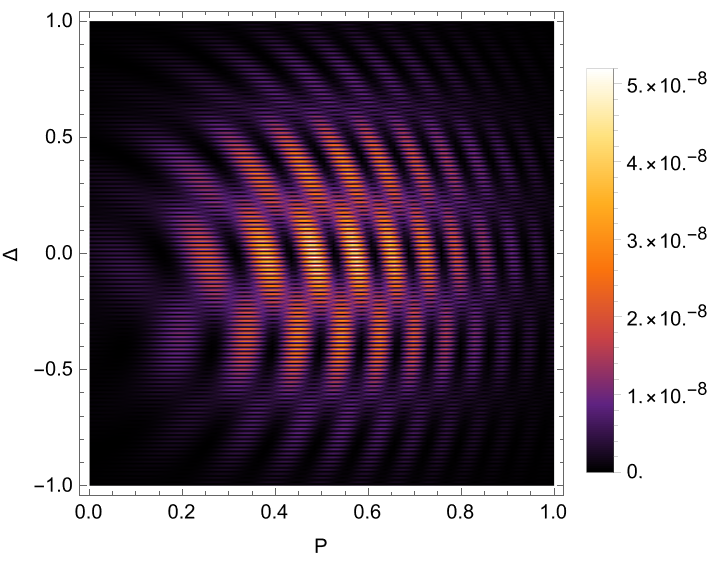}
    \caption{Spectrum for~\eqref{fourPulses} with $\gamma=1$ and $E_0=0.2$ showing moir\'e patterns. The four pulses are placed as in~\eqref{fourPulsesPositionsScattering}. The first row is obtained with only the $a$ and $b$ pulses, the second with only the $c$ and $d$ pulses, and the third row with all four pulses. The left column is obtained with the grid approach, and the right column is obtained with the quadratic approximation.}
    \label{moire4scattering}
\end{figure*}

\begin{figure}
    \centering
    \includegraphics[width=\linewidth]{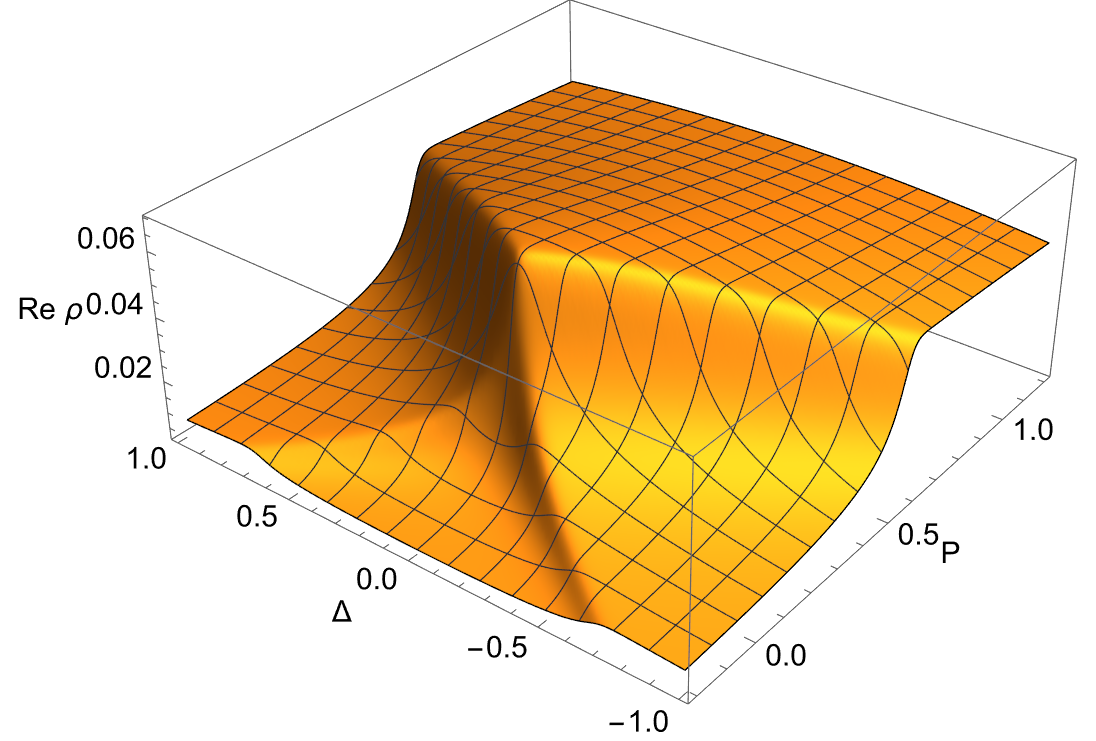}
    \caption{Same notation as in Fig.~\ref{fig:actuallySlow}, but for pulse $a$ in~\eqref{fourPulsesPositionsScattering}.}
    \label{fig:rePre3DLargez}
\end{figure}

\begin{figure*}
    \includegraphics[width=.49\linewidth]{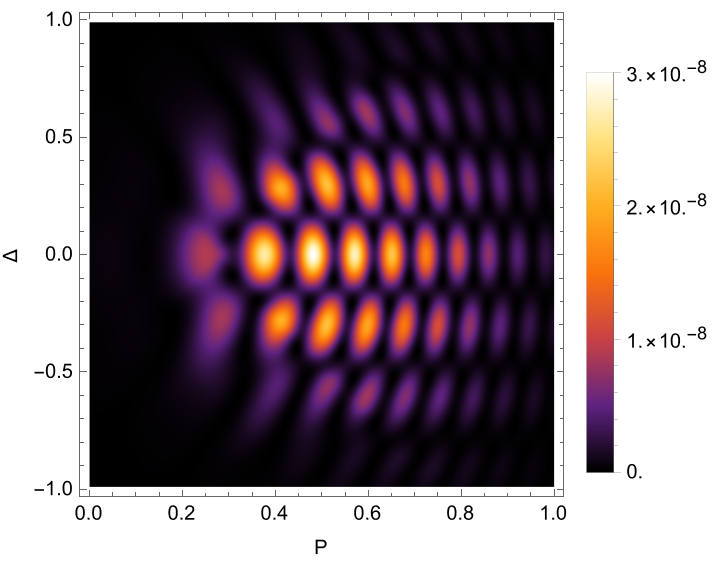}
    \includegraphics[width=.49\linewidth]{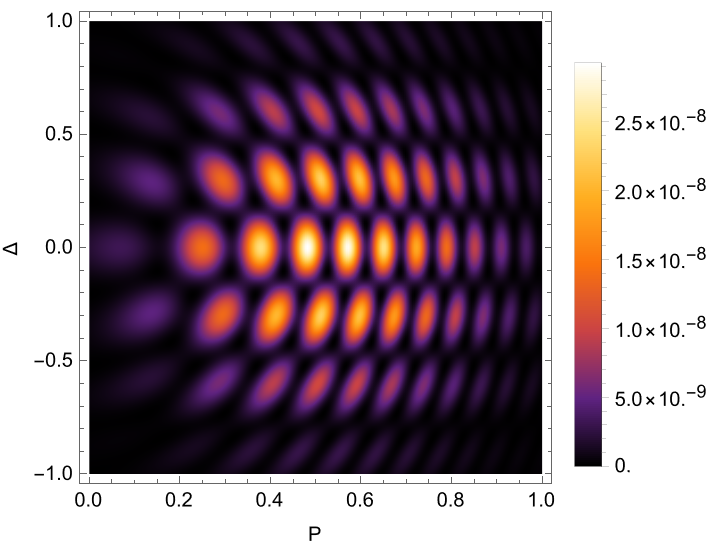}
    \caption{Spectrum for three pulses~\eqref{threePulses} placed as in~\eqref{threePulsesScattering}, with $\gamma=1$ and $E_0=0.2$. The first plot shows the full/grid-approach result, and the second plot shows the quadratic approximation.}
    \label{fig:moire3t}
\end{figure*}

In the first moir\'e-pattern example~\eqref{fourPulsesPositions}, the peaks are placed such that each instanton only goes through one peak. To illustrate what can happen if this is not the case, we consider 
\be\label{fourPulsesPositionsScattering}
\begin{split}
(t_a,z_a)&=(0,0) \quad \gamma(t_b,z_b)=(15,0) \\
\gamma(t_c,z_c)&=(0,70) \quad \gamma(t_d,z_d)=(15,73) \;.
\end{split}
\ee
An instanton created at pulse $a$ ($c$) can, for certain values of $\Pi$, scatter on pulse $b$ ($d$). We compare the results from the grid approach and the quadratic approximation in Fig.~\ref{moire4scattering}. The two methods lead to similar results, but there are noticeable differences in some regions of momentum space. This is to be expected. Consider the asymptotic straight-line paths of the instanton created in pulse $a$,
\be
q_{(a)}^\mu(u)~\sim q_a^\mu+q_{(a)}^{\prime\mu}(\pm\infty)u \;.
\ee
This instanton will see pulse $b$ if $z_{(a)}(t_b)\sim z_b$, i.e. if
\be
z_b~\sim z_a+\frac{z_{(a)}'(\pm\infty)}{t_{(a)}'(\pm\infty)}(t_b-t_a) \;.
\ee
If the two pulses are only separated in time, $z_a=z_b$, we have scattering if either the electron or the positron emerges from pulse $a$ with $z'\sim0$, i.e.
\be
\Delta\sim\pm 2P \;.
\ee
Within the wedge $|\Delta|\lesssim 2P$ there is no scattering and the quadratic approximation works well, as can be seen in the first two plots in Fig.~\ref{moire4scattering}. This is good news because the momentum saddle point, and hence also the numerically most important region, lie in this wedge. We can see some differences around the lines $\Delta\sim\pm 2P$. The difference can be most clearly seen in a plot of the prefactor for the amplitude term associated with instanton $a$. Indeed, if we compare Fig.~\ref{fig:rePre3DLargez} with the corresponding result in the case without scattering, Fig.~\ref{fig:actuallySlow}, we see that something dramatic happens near $\Delta\sim\pm 2P$, and also outside the wedge.

As a second example, we consider three pulses~\eqref{threePulses} placed at
\be\label{threePulsesScattering}
(t_a,z_a)=(0,0)
\quad
\gamma(t_b,z_b)=(15,2)
\quad
\gamma(t_c,z_c)=(0,4) \;.
\ee
The resulting spectrum is shown in Fig.~\ref{fig:moire3t}. Here both the $a$ and the $c$ pulses will, in some parts of momentum space, scatter on pulse $b$. 

As mentioned, we use the shooting method to find the instanton at each grid point. For this we need a good starting point for $t(0)$ and $z(0)$. In the case without scattering, we can simply use the values of $t(0)$ and $z(0)$ for $\Pi=\Pi_s$ as starting point for finding the instanton for $\Pi\ne\Pi_s$. But this approach often breaks down in the case with scattering, because even a small change in $\Pi$ might lead to a discontinuous and incorrect jump in $q^\mu(0)$. This is not surprising because if pulse $b$ ``subtends a small solid angle''\footnote{It is a bit more complicated than that, since the instanton can be complex.} as seen by pulse $a$, then even a very small change in the momentum immediately after emerging from pulse $a$ can mean the difference between scattering and no scattering. To solve this issue we have used a numerical continuation, where we start with the instanton for $\Pi=\Pi_s$ and take small steps along the $P$ direction, 
\be
\begin{split}
P&=P_s, P_s-\delta P, P_s-2\delta P, \dots, 0 \\
P&=P_s, P_s+\delta P, P_s+2\delta P, \dots, 1 \;,
\end{split}
\ee
with $\Delta=\Delta_s$, and at each step we use the values of $t(0)$ and $z(0)$ from the previous step as starting point. Then we use the result on the $P$ axis to make a similar continuation in the $\Delta$ direction.    

\subsection{Additional instantons?}

\begin{figure*}
    \centering
    \includegraphics[width=\linewidth]{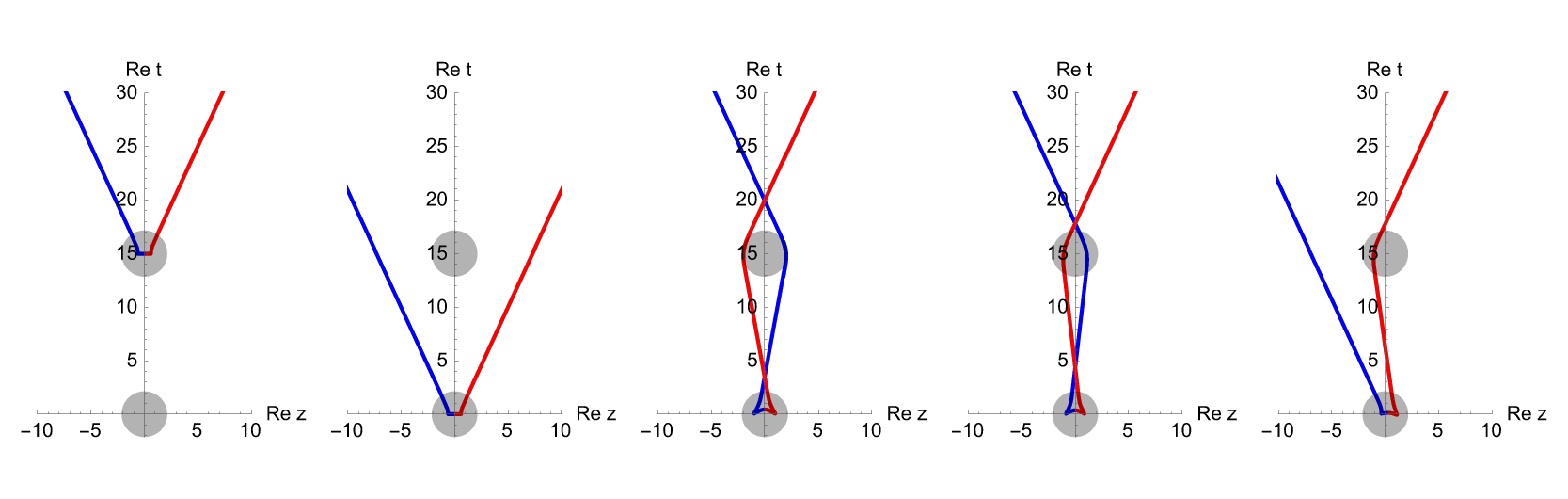}\\
    \includegraphics[width=\linewidth]{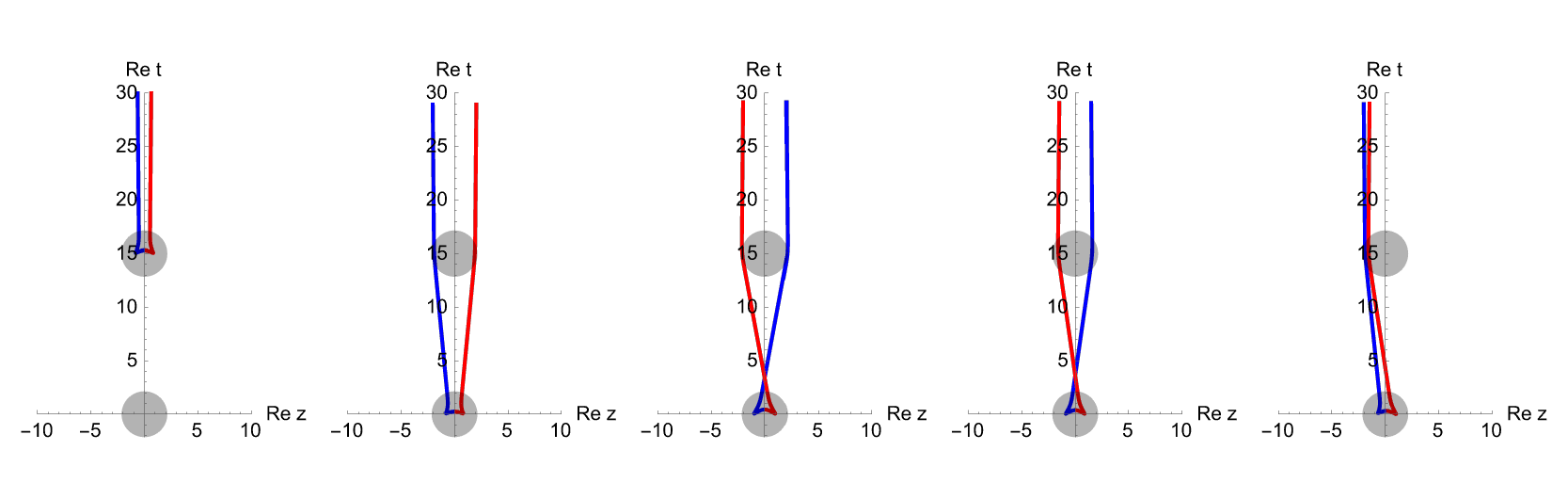}\\
    \includegraphics[width=\linewidth]{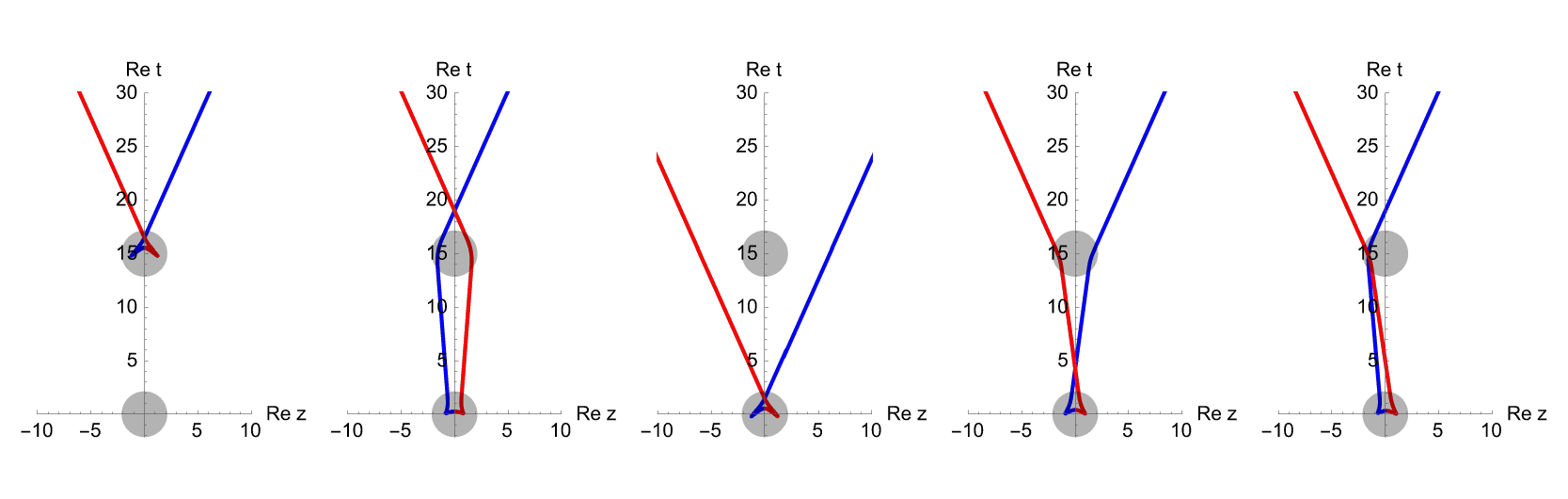}
    \caption{Real part of instantons for two Gaussian pulses, with $\gamma_t=\gamma_z=1$, located around the gray disks. The radius $r$ of the disks is chosen such that $e^{-r^2}=0.01$. The asymptotic momentum is $P=P_s\sim0.52$ in the first row, $P\sim0$ in the second, $P=-0.5$ in the third, and $\Delta=0$ in all plots. The instantons within one column are continuously connected. The red (blue) part of the instanton corresponds to the electron (positron). There is an additional class of instantons which are the mirror image of the asymmetric instantons.}
    \label{fig:additionalInstantons}
\end{figure*}

The instantons we considered in the previous section are all continuously connected to the instantons which give a peak in the spectrum at $\Pi=\Pi_s\sim0.52$ and $\Delta=\Delta_s=0$ for $\gamma_t=\gamma_z=1$. By comparing Fig.~\ref{fig:rePre3DLargez} and Fig.~\ref{fig:actuallySlow} we see that something dramatic happens at the wedge $\Delta\sim\pm2P$. But this is not surprising, because this is simply due to the fact that pulse $b$ scatters instanton $a$ in that region of momentum space, where $\Delta\sim\pm2P$. However, Fig.~\ref{fig:rePre3DLargez} and Fig.~\ref{fig:actuallySlow} also show that there continues to be a large difference even far outside the wedge. In other words, if we start at $P=P_s\sim0.52$ and $\Delta=\Delta_s=0$ and make a numerical continuation to either larger $\Delta$ or negative $P$, then at $\Delta\sim\pm2P$ $q_{(a)}^\mu$ starts to scatter against pulse $b$, but if we continue past this wedge $q_{(a)}^\mu$ still continues to be scattered by pulse $b$. This is illustrated in the second column in Fig.~\ref{fig:additionalInstantons}. A dense set of $q_{(a)}^\mu$ instantons on a grid in $\Pi$ gives Fig.~\ref{fig:rePre3DLargez}. The first column shows three snapshots of a numerical continuation of the instanton which we call $q_{(b)}^\mu$.
$q_{(a)}^\mu$ and $q_{(b)}^\mu$ give the interference patterns shown in the previous sections.  

Compare the first three plots in the third row, where the asymptotic momentum is $P=-0.5$ and $\Delta=0$. One might naively have guessed that this point would lie sufficiently far from the wedge $\Delta\sim\pm2P$ that there should no longer be any scattering for $q_{(a)}^\mu$. But we see that $q_{(a)}^\mu$ still goes through pulse $b$. If one just takes $q_{(b)}^\mu$ from the plot and guess what the corresponding instanton for $q_{(a)}^\mu$ would be, then one might have guessed the instanton in the third plot. This instanton, call it $q_{(3)}^\mu$, seems perfectly physical, at least in this part of momentum space. But if we increase $P$ again up to $P\sim0.52$, then $q_{(3)}^\mu$ turns into the instanton in the first row and third column, which goes through pulse $b$ even within the wedge, i.e. in a region where one, at least at first, would not have expected any scattering. 

The class of instantons in the fourth column, $q_{(4)}^\mu$, seem quite similar to $q_{(3)}^\mu$ for $P>0$, but for $P=-0.5$ $q_{(4)}^\mu$ still goes through pulse $b$ while $q_{(3)}^\mu$ no longer feels pulse $b$. The last column shows an asymmetric class of instantons, $q_{(5)}^\mu$, which nevertheless have the same asymptotic momenta as the other instantons. There are also instantons which are mirror image of $q_{(5)}^\mu$.  

Now, just because there are additional instantons (saddle points) does not necessarily mean that one should include them. Whether or not to include a saddle point is in principle determined by the ``intersection numbers'' in the Lefschetz-thimble formalism, see e.g.~\cite{Aniceto:2018bis}. However, that is beyond the scope of this paper. We note instead the following problem with these additional instantons. Of the instantons in Fig.~\ref{fig:additionalInstantons} which are created by pulse $a$, at least in the plotted region, only $q_{(a)}^\mu$ gives a contribution with a saddle point for $\Pi$. In other words, the gradient of the real part of the exponent, 
\be
\nabla\text{Re}\psi=\text{Re}(\partial_P,\partial_\Delta)\psi \;,
\ee
vanishes at $\Pi=\Pi_s$ for $q_{(a)}^\mu$, but for the other instantons $\nabla\text{Re}\psi$ does not vanish at $\Pi_s$ or in any other region we have considered. We have found that if we start at $\Pi_s$ and numerically continue $q_{(j)}^\mu$ along the line in $\Pi$ space which is everywhere parallel to the gradient, then $\text{Re}\psi[q_{(j)}]$ increases and eventually switches sign, making $\exp(\psi[q_{(j)}]/E)\gg1$ instead of $\exp(\psi[q_{(j)}]/E)\ll1$. An exponentially large term is obviously unphysical. So, at least in a region around the point where the exponent switches sign, such a class of instantons is unphysical and should not be included.

However, this does not rule out the possibility that a class of instantons can be unphysical in some region of momentum space, but physical in another region, as a type of Stokes phenomenon. In particular, if we take the limit where the field strength of pulse $b$ goes to zero, $E_b\to0$, then $q_{(3)}$ in the third column, last row in Fig.~\ref{fig:additionalInstantons} would not change (since pulse $b$ is already negligible on that instanton path even before taking $E_b\to0$). The same instanton would also be obtained if one started with $E_b=0$ identically and numerically continued $q_{(a)}$ from $\Pi_s$ to $P=-0.5$, $\Delta=0$. This suggests that $q_{(3)}$ is physical also in the presence of pulse $b$ in the region around that momentum point. 

Regardless of whether or not there are additional or different instantons in some regions of momentum space, we can still use the methods we have developed here. We leave a more detailed investigation of additional instantons for future studies.

\section{Eigenvalues}\label{sec:eigenvalues}

Expanding the exponential part of the integrand of the worldline path integral to second order around the instanton gives a Gaussian path integral with the following exponent,
\be
\exp\left\{-\frac{i}{2}\int\ud u \, \delta q^\mu\Lambda_{\mu\nu}\delta q^\nu\right\} \;,
\ee
where $\delta q^\mu=q^\mu-q_\text{saddle}^\mu$. With
\be
{\bm\Lambda}=\begin{pmatrix}{\Lambda^0}_0&{\Lambda^0}_3\\ {\Lambda^3}_0&{\Lambda^3}_3\end{pmatrix}
\ee
we have
\be
{\bm\Lambda}=\begin{pmatrix}
-\partial_u^2+E_{,t}z' & E_{,z}z'+E\partial_u\\
E_{,t}t'+E\partial_u & -\partial_u^2+E_{,z}t'
\end{pmatrix} \;,
\ee
where $E_{,t}=\partial_t E$ etc.
We have previously focused on the determinant of $\Lambda$. Now we will consider its eigenvalues 
\be\label{eq:eigenEq}
{\bm\Lambda} \, {\bm\phi}(u)=k^2{\bm\phi}(u)
\ee
with boundary conditions $\bm\phi(u_0) = \bm\phi(u_1) = 0$. The eigenvalues, and thus the determinant of the operator $\bm \Lambda$, depend on the boundary conditions $u_0$ and $u_1$ but not on the choice of contour between them. For example, in the free case we have
\be\label{k0free}
k_0 = \frac{n \pi}{u_1 - u_0} \; .
\ee
The distribution of eigenvalues becomes denser and denser as the length of the proper-time interval, $u_1-u_0$, increases. This can make it challenging to obtain a sufficient number of eigenvalues for non-constant fields. The following approach helps. 

We begin by transforming~\eqref{eq:eigenEq} to a Liouville normal form, which we achieve by writing $\phi$ in terms of $\varphi$ defined by
\be
{\bm\phi}(u)=\begin{pmatrix}
    \cosh\left(\frac{\nu}{2}\right)&\sinh\left(\frac{\nu}{2}\right)\\
    \sinh\left(\frac{\nu}{2}\right)&\cosh\left(\frac{\nu}{2}\right)
\end{pmatrix}
{\bm\varphi}(u) \;,
\ee
where
\be
\nu(u)=\int_{u_0}^u\ud v\, E[t(v),z(v)] \;,
\ee
which gives
\be
(\partial_u^2+k^2-{\bf Q}){\bm\varphi}=0 \;,
\ee
where 
\be
\begin{split}
2{\bf Q}&=\frac{E^2}{2}+\nabla E\cdot\{z',t'\}\\
&+(E_{,t}z'-E_{,z}t')\begin{pmatrix}
    \cosh\,\nu &\sinh\,\nu\\
    -\sinh\,\nu &-\cosh\,\nu
\end{pmatrix}\\
&+(E_{,z}z'-E_{,t}t')\begin{pmatrix}
    \sinh\,\nu &\cosh\,\nu\\
    -\cosh\,\nu &-\sinh\,\nu
\end{pmatrix} \;.
\end{split}
\ee
We can perform the integral in $\nu$ using
\be
E[t(u),z(u)]=\pm\partial_u\ln[\pm t'+z'] \;,
\ee
and we can simplify the trig functions using
\be
\begin{split}
\cosh[\ln b-\ln a]&=\frac{1}{2}\left(\frac{b}{a}+\frac{a}{b}\right)
\\
\sinh[\ln b-\ln a]&=\frac{1}{2}\left(\frac{b}{a}-\frac{a}{b}\right) \;.
\end{split}
\ee

We can find several of the first eigenvalues numerically, but since we are eventually interested in multiplying all eigenvalues, it is useful to derive analytical approximations for large eigenvalues. One way to do this is to first rewrite the eigenvalue equation as a Volterra integral equation, as described in e.g.~\cite{CourantHilbert,Carlson},
\be
{\bm\varphi}(u)={\bf A}\sin[k(u-u_0)]+\frac{1}{k}\int_{u_0}^u\ud v\,\sin[k(u-v)]{\bf Q}(v){\bm\varphi}(v) \;,
\ee
where ${\bf A}=(A_1,A_2)$ is a constant vector and we have imposed the Dirichlet boundary condition at $u=u_0$. The eigenvalue $k^2$ is determined by imposing the Dirichlet condition at $u=u_1$. For large $k$ we can, to next-to-leading order (NLO), replace ${\bm\varphi}\to {\bf A}\sin(...)$ in the $v$ integral,
\be
\begin{split}
{\bm\varphi}(u_1)&\approx\bigg\{\sin[k\Delta u]\\
&+\frac{1}{k}\int_{u_0}^{u_1}\ud v\sin[k(u_1-v)]\sin[k(v-u_0)]{\bf Q}\bigg\}{\bf A}\\
&\approx\bigg\{\sin[k\Delta u]-\frac{\cos[k\Delta u]}{2k}\int_{u_0}^{u_1}\ud v\,{\bf Q}(v)\bigg\}{\bf A} \;,
\end{split}
\ee
where $\Delta u=u_1-u_0$ and where we have dropped terms in the integrand which oscillate rapidly for large $k$.
For ${\bm\varphi}(u_1)=0$ to have nontrivial solutions, the determinant of the matrix multiplying the constant ${\bf A}$ must vanish,
\be\label{det0vers1}
\det\bigg\{\sin[k\Delta u]-\frac{\cos[k\Delta u]}{2k}\int{\bf Q}\bigg\}=0 \;.
\ee
To zeroth order we have $k_0=n\pi/\Delta u$, where $n$ is an integer. Note that $k_0$ is independent of the field, so it agrees in particular with the eigenvalues of the free path integral~\eqref{k0free}. To NLO, $k\approx k_0+k_1$, we have
\be\label{det0vers2}
\det\left(k_1-\frac{1}{2\pi n}\int{\bf Q}\right)=0 \;,
\ee
which has two solutions for each $n$,
\be\label{eigenvalueNLO}
k_1=\frac{1}{2\pi n}\left(\frac{1}{2}\text{tr}\int{\bf Q}\pm\sqrt{\left(\frac{1}{2}\text{tr}\int{\bf Q}\right)^2-\det\int{\bf Q}}\right) \;.
\ee

Thus, $k_1/k_0$ is proportional to $1/n^2\ll1$. But $k_1/k_0$ is also proportional to $\Delta u$, and, for example, to go from~\eqref{det0vers1} to~\eqref{det0vers2} we need to assume $k_1\Delta u\ll1$, i.e. $n\gg\Delta u$. This is what can make it challenging to obtain enough eigenvalues, because for our open worldlines the proper-time interval $\Delta u=u_1-u_0$ needs to be sufficiently large so that the instanton is outside the field when $\text{Re }u<\text{Re }u_0$ and $\text{Re }u>\text{Re }u_1$. But the larger $\Delta u$, the larger $n$ has to be for the eigenvalues to start settling down to their asymptotic form. And we need them to settle down so that we can extrapolate the product of $\prod_{n=1}^N\sqrt{\lambda_n}/\sqrt{\lambda_n^{(0)}}$ to its $N\to\infty$ limit. 

As a check, for a constant field the square root in~\eqref{eigenvalueNLO} vanishes, so the two eigenvalues for each $n$ stay degenerate,
\be
k_1=\frac{\Delta u}{8\pi n} \;,
\ee
which agrees with the expansion of the exact eigenvalues
\be
k=\left(\left[\frac{n\pi}{\Delta u}\right]^2+\frac{1}{4}\right)^{1/2} \;.
\ee

Having found an asymptotic approximation of the eigenvalues~\eqref{eigenvalueNLO} allows us to check that the numerically computed eigenvalues converge correctly. It also means that we can improve the precision of the product of the eigenvalues beyond what a finite number of them gives. To obtain a convergent product we follow the usual procedure and divide by the product of the free eigenvalues,
\be\label{eq:Peq}
P=\prod\limits_{n=1}^\infty\frac{\sqrt{\lambda_n}}{\sqrt{\lambda^{(0)}_n}} \;.
\ee
If we have computed the first $m-1$ eigenvalues numerically, then we have
\be
P=P_{n<m}P_{n\geq m} \;,
\ee
where the precision of $P_{n<m}$ is determined by the precision of the numerically obtained eigenvalues, and $P_{n\geq m}\approx1$ for sufficiently large $m$. But we can do better by approximating $P_{n\geq m}$
using~\eqref{eigenvalueNLO}. If we imagine going beyond the NLO~\eqref{eigenvalueNLO}, we would have
\be\label{Pnmab}
P_{n\geq m}=\prod\limits_{n=m}^\infty\left(1+\frac{a_1}{n^2}+\frac{a_2}{n^4}+\dots\right)=1+\frac{b_1}{m}+\frac{b_2}{m^2}+\dots \;,
\ee
where $a_1$ is obtained from~\eqref{eigenvalueNLO} and $a_{n>1}$ could be obtained by expanding the Volterra equation to higher orders. The coefficients $b_k$ can be obtained by expanding 
\be
\begin{split}
0=&\left(1+\frac{a_1}{m^2}+\frac{a_2}{m^4}+\dots\right)^{-1}P_{n\geq m}-P_{n\geq m+1}\\
=&\left(1+\frac{a_1}{m^2}+\frac{a_2}{m^4}+\dots\right)^{-1}\left(1+\frac{b_1}{m}+\frac{b_2}{m^2}+\dots\right)\\
&-\left(1+\frac{b_1}{m+1}+\frac{b_2}{(m+1)^2}+\dots\right)\\
=&b_1-a_1+\frac{2b_2-(1+a_1)b_1}{m}+\dots
\end{split}
\ee
and demanding that each order in $1/m$ vanish,
which gives for the first three coefficients
\be
b_1=a_1
\quad
b_2=\frac{a_1(1+a_1)}{2}
\quad
b_3=\frac{a_1(1+a_1)^2}{6}+\frac{a_2}{3} \;.
\ee
Note that $a_2$ only enters at $\mathcal{O}(1/m^3)$, so for $\mathcal{O}(1/m)$ and $\mathcal{O}(1/m^2)$ we only need $a_1$, i.e. the coefficient we have already calculated in~\eqref{eigenvalueNLO}. 

For the free case, each eigenvalue, $k=n\pi/\Delta u$, is doubly degenerate, while the correction in~\eqref{eigenvalueNLO} splits this into two different values. This gives two different $a_1$ coefficients
\be
a_1^\LCpm=\frac{\Delta u}{2\pi^2}\left(\frac{1}{2}\text{tr}\int{\bf Q}\pm\sqrt{\left(\frac{1}{2}\text{tr}\int{\bf Q}\right)^2-\det\int{\bf Q}}\right) \;.
\ee
For the leading order (LO) correction the square root drops out,
\be
\left[1+\frac{a_\LCp}{n^2}\right]\left[1+\frac{a_\LCm}{n^2}\right]=1+\frac{a'_1}{n^2}+\mathcal{O}(n^{-4}) \;,
\ee
where
\be
\begin{split}
a'_1&=a_1^\LCp+a_1^\LCm=\frac{\Delta u}{2\pi^2}\text{tr}\int{\bf Q}\\
&=\frac{\Delta u}{2\pi^2}\int_{u_0}^{u_1}\ud v\left(\frac{E^2}{2}+\nabla E\cdot\{z',t'\}\right) \;.
\end{split}
\ee
Thus,
\be\label{eq:Pcorrections}
P_{n\geq m}=1+\frac{a'_1}{m}+\frac{a'_1(1+a'_1)}{2m^2}+\mathcal{O}(m^{-3}) \;.
\ee

\begin{figure}
    \includegraphics[width=\linewidth]{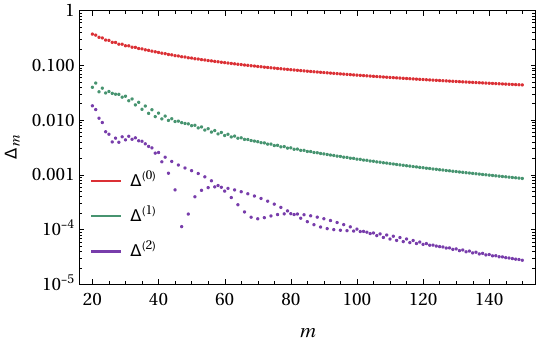}
    \caption{Comparisons~\eqref{eq:deltas} for one of the two instantons of the field~\eqref{eq:field21} with $\gamma_t = \gamma_z = 1$ using a contour $u(r)=e^{-0.1\pi i}r$ with $-10<r<10$.}
    \label{fig:eigsComparison}
\end{figure}

In Fig.~\eqref{fig:eigsComparison}, we compare the result obtained using the Gelfand-Yaglom method with $P_{n<m}$, $P_{n<m}$ and the first-order correction, and $P_{n<m}$ including both the first- and second-order corrections~\eqref{eq:Pcorrections}, i.e.
\be\label{eq:deltas}
\begin{split}
    \Delta^{(0)}_m &= \left|\frac{P_{n<m}}{P} - 1 \right| \\
    \Delta^{(1)}_m &= \left|\frac{P_{n<m}\left( 1+\frac{a'_1}{m} \right)}{P} - 1 \right| \\
    \Delta^{(2)}_m &= \left|\frac{P_{n<m}\left( 1+\frac{a'_1}{m}+\frac{a'_1(1+a'_1)}{2m^2} \right)}{P} - 1 \right| \;.
\end{split}
\ee
We have calculated the exact eigenvalues (which give $P_{n<m}$) using the built-in function ``NDEigenvalues'' in Mathematica.
We see that even if we include $m>100$ of the first eigenvalues, the relative error for $P_{n<m}$ is quite large. However, by including the correcting factor, the error is reduced by orders of magnitude.

\subsection{Morse-Maslov index}

As we explicitly demonstrated in the previous section, in the cases we consider the eigenvalues are complex. To compare our treatment with the literature, let us first recall the following. 
For differential operators $\Lambda$ with real eigenvalues, there is an additional phase $e^{-i\nu\pi/2}$ at the prefactor~\cite{Dunne:2006st} called the Morse-Maslov index~\cite{Levit:1976fv,Morse} arising when one goes from an infinite product of square roots~\eqref{eq:Peq} to the square root of an infinite product, as it appears in the Gelfand-Yaglom method. The extra phase, $e^{-i\nu\pi/2}$, with $\nu$ the number of negative eigenvalues, fixes this. For example, if one has only two negative eigenvalues, the product of square roots~\eqref{eq:Peq} has an extra factor of $i^2=-1$, but if all the eigenvalues are under the same square root the signs cancel out. In other words, the Morse-Maslov index is the total accumulated phase in the product~\eqref{eq:Peq}. A more convenient method to calculate this contribution is using the Morse index theorem~\cite{Morse}, which states that the number of negative eigenvalues is equal to the number of times the quantity $\det \Lambda(\tau)$ vanishes in the interval $(0,1)$. If the path integral~\eqref{eq:PIsaddles} only receives contributions from one instantons, one does not need to calculate it explicitly because it just amounts to an overall phase, which is irrelevant since with our open worldlines we are working on the amplitude rather than the probability level. However, in this work, since we consider fields where multiple instantons contribute, we need to take this into account.

For open instantons, the eigenvalues are in general complex, and so is $\det \Lambda(\tau)$, so we cannot use the standard Morse index theorem. The problem of fixing the phase amounts to choosing what branch of the square root of
\be\label{sqrtProd}
\sqrt{\prod_{n=1}^\infty \frac{\lambda_n}{\lambda^{(0)}_n}} = \sqrt{\frac{\det \Lambda(u_1)}{\det \Lambda_0(u_1)}} = \sqrt{\frac{t_\LCp t_\LCm}{T p_0 p'_0} h(u_1)}
\ee
one has to take. To do this, we look at the accumulated phase of the product of the eigenvalues. For the fields we consider, we find that it is always between $(-\pi, \pi)$, therefore when taking the square root of the right-hand side we take the phase to lie in this interval. In the hypothetical case above with two eigenvalues, the accumulated phase is $2\pi$, therefore one must choose the branch $(e^{2\pi i})^{1/2} = e^{i \pi} = -1$.
Note, though, if we just consider~\eqref{eq:Peq} instead of~\eqref{sqrtProd}, then we automatically obtain the correct phase without having to fix anything or keeping track of some accumulated phase. However, that does not mean that we want to abandon the Gelfand-Yaglom approach, because: 1) It is of course always good to have two different approaches to calculate the same thing. 2) While we can obtain the determinant for moderately large $\Delta u$ with either approach, the GY approach is faster. 3) In~\cite{DegliEsposti:2022yqw} we showed how to take the $\Delta u\to\infty$ limit analytically within the GY approach, which significantly simplifies the remaining numerical computations.

\section{Conclusions}

We have shown how to use worldline instantons to obtain the spectrum of Schwinger pair production for spacetime dependent electric fields which have multiple peaks, which leads to interference patterns in the momentum spectrum. We have found that not only does the spatial dependence affect the patterns, the conditions for when there is any interference at all changes completely compared to the purely time-dependent-field case.

For $E(t)$ there is momentum conservation which means that the sum of the electron and positron momentum is zero, $p_3+p_3'=0$. This is no longer true for $E(t,z)$. While fields of the form $E(t,z)$ have been studied before, as far as we are aware, only the dependence on either $p_3$ or $p_3'$ has been studied, but not $p_3$ and $p_3'$ together. We have found 2D interference patterns in the $p_3$-$p_3'$ plane such as moir\'e patterns.

Our general approach works for rather general field configurations, but we have also derived some simpler approximations for the cases where the field peaks are well separated. We observed that the frequency of oscillations in the spectrum diverges in the LCF limit, which means that the spectrum basically becomes an incoherent sum of contributions from individual peaks. Away from the LCF limit we have seen rich interference patterns. But it would be interesting to study what sources of decoherence might be relevant and under what conditions.

While we have focused in this paper on fields which only depend on one spatial coordinate, our previous results~\cite{DegliEsposti:2023qqu} for fields with a single peak (and hence no interference) suggests that it should be straightforward to generalize also the results in the present paper to fully 4D fields. 

Moreover, while we have focused on Schwinger pair production, which is not the most experimentally relevant process, based on~\cite{DegliEsposti:2023fbv} we expect that one should be able to generalize to other, more experimentally relevant strong-field processes such as nonlinear Breit-Wheeler pair production.

\acknowledgements

G. T. is supported by the Swedish Research Council, Contract No. 2020-04327.

\appendix

\section{Spin factor}\label{App:Spin}

We deal with the spin factor as in~\cite{DegliEsposti:2022yqw}. Denoting with $E(u)$ the electric field evaluated at the instantons and using
\be
E(u) = \frac{\ud}{\ud u} \ln(t'(u) + z'(u))
\ee
we deal with the integral at the exponent of the spin factor by splitting it into the $u>0$ and $u<0$ regions. Using~\eqref{eq:SPE} (and the analogous equation for $t$), $t'(0) = 0$ and $z'(0) = \varepsilon i$ we obtain
\be
\begin{split}
    \frac{1}{2}\int \ud u \, E(u)
    &= \frac{1}{2} \ln \frac{p_0 + p_3}{z'(0)} + \frac{1}{2} \ln \frac{-z'(0)}{p'_0 - p'_3} \\
    &= \frac{1}{2} \ln \frac{p_0 + p_3}{p'_0 - p'_3} - \ln \varepsilon i \\
    &= \frac{1}{2} \ln \frac{p_0 + p_3}{p'_0 - p'_3} -\frac{i \varepsilon \pi}{2} \; .
\end{split}
\ee
The first term is treated as in~\cite{DegliEsposti:2022yqw}, while the second term, coming from the sign of $z'(0)$, is a phase that gives the relative sign in~\eqref{eq:Mj}.

\section{Symmetric fields}\label{App:Symmetric}

If we have an electric field $E_3(t,z)$ which is even with respect to $z$, i.e. $E_3(t,z) = E_3(t,-z)$, we find that the saddle-point instantons have an even time component $t(u)$ and an odd space component $z(u)$. In fact, the Lorentz-force equation~\eqref{LorentzEq} is consistent with said symmetries regardless of the time-symmetry. In this case the expressions for the coefficients~\eqref{eq:GeneralParameters} are much simpler and some of them are zero. First of all, due to the symmetry of the instantons, we see that $\Delta_s = 0$ and $\alpha_\Delta = 0$. Furthermore, the variations $\delta t_P$ and $\delta z_\Delta$ are even while $\delta t_\Delta$ and $\delta z_P$ are odd, which means that $X_{\Delta P} = 0$ hence $d^{-2}_{\Delta P} = \beta_{\Delta P} =0$. 

Focusing on the longitudinal part of the spectrum ($p_\LCperp =p'_\LCperp = 0$) and letting $\alpha = \alpha_P$, the exponent simplifies to
\be
\begin{split}
-\frac{\mathcal A}{2}& + i\phi + i\alpha(P -P_s) + \frac{1}{2}\left( -\frac{1}{d_P^2} +i\beta_P \right) (P -P_s)^2 \\
&+ \frac{1}{2}\left(-\frac{1}{d_\Delta^2}+i\beta_\Delta \right) \Delta^2 \;,
\end{split}
\ee
where the widths have the same expressions as in~\cite{DegliEsposti:2022yqw} and the remaining constants are obtained by substituting the imaginary part for the real part, namely
\be
\begin{split}
    \{ \phi,\mathcal A/2 \} &= \{\Re,\Im\} \, \int \ud u \, q^\mu \pa_\mu A_\nu \, \frac{\ud q^\nu}{\ud u} \\
    \alpha &= 2\Re \left[\frac{P_s}{p_0}t(\infty) -z(\infty)\right] \\
    \{\beta_P, d_P^{-2}\} &= \frac{2}{p_0^2} \{\Re,\Im\} \left[ \frac{t(\infty)}{p_0} -\frac{\eta_{s}(\infty)}{\eta'_{s}(\infty)}\right] \\
    \{\beta_\Delta, d_\Delta^{-2}\} &= \frac{1}{2p_0^2} \{\Re,\Im\} \left[ \frac{t(\infty)}{p_0} -\frac{\eta_{a}(\infty)}{\eta'_{a}(\infty)}\right]
\end{split}
\ee
where $\eta_{a,s}$ are respectively antisymmetric and symmetric solutions to 
\be
\eta'' = (E^2 + \nabla E \cdot\{z',t'\}) \eta
\ee
with initial conditions
\be
\begin{split}
    \eta_s(0) &= 1 \qquad \eta_a(0) = 0 \\
    \eta'_s(0) &= 0 \qquad \eta'_a(0) = 1  \; .
\end{split}
\ee
The spectra for these fields is qualitatively different from the ones of nonsymmetric fields because $\alpha_\Delta = 0$ for each term in~\eqref{eq:specTotal} implies that the interference pattern is only in the $P$ direction, as can be seen in the first plot in Fig.~\ref{fig:spectra1}.

\end{document}